\documentclass[twocolumn,superscriptaddress,amssymb,amsmath,nobibnotes, aps, prab,10pt]{revtex4-1}
\expandafter\ifx\csname package@font\endcsname\relax\else
\expandafter\expandafter
\expandafter\usepackage
\expandafter\expandafter
\expandafter{\csname package@font\endcsname}%
\fi
\usepackage{hyperref}
\usepackage{graphicx}
\usepackage{booktabs}
\usepackage{amssymb}
\usepackage{amsfonts}
\usepackage{amsmath}
\usepackage[justification=justified,format=plain]{caption}
\usepackage{float}
\usepackage{textcomp}
\usepackage{epstopdf}
\usepackage{gensymb}
\usepackage{blindtext}
\usepackage{siunitx}
\usepackage{tabularx}
\usepackage{enumitem}
\usepackage{subfig}
\usepackage{fourier} 
\usepackage{makecell}
\usepackage{tikz}
\usepackage[margin=0.5in]{geometry}

\hypersetup{pdfnewwindow}

\begin{document}
\title{Microbunching Instability Characterisation via Temporally Modulated Laser Pulses}
\author{A.~D.~Brynes}
\email[]{alexander.brynes@stfc.ac.uk}
\affiliation{ASTeC, STFC Daresbury Laboratory, Daresbury, Warrington, WA4 4AD Cheshire, United Kingdom}
\affiliation{Cockcroft Institute, Sci-Tech Daresbury, Keckwick Lane, Daresbury, Warrington, WA4 4AD, United Kingdom}
\affiliation{Department of Physics, University of Liverpool, Liverpool, L69 7ZE, United Kingdom}
\author{I.~Akkermans}
\affiliation{ASML Netherlands B.V., 5504 DR Veldhoven, Netherlands}
\author{E.~Allaria}
\affiliation{Elettra-Sincrotrone Trieste S.C.p.A., 34149 Basovizza, Trieste, Italy}
\author{L.~Badano}
\affiliation{Elettra-Sincrotrone Trieste S.C.p.A., 34149 Basovizza, Trieste, Italy}
\author{S.~Brussaard}
\affiliation{ASML Netherlands B.V., 5504 DR Veldhoven, Netherlands}
\author{M.~Danailov}
\affiliation{Elettra-Sincrotrone Trieste S.C.p.A., 34149 Basovizza, Trieste, Italy}
\author{A.~Demidovich}
\affiliation{Elettra-Sincrotrone Trieste S.C.p.A., 34149 Basovizza, Trieste, Italy}
\author{G.~De~Ninno}
\affiliation{Elettra-Sincrotrone Trieste S.C.p.A., 34149 Basovizza, Trieste, Italy}
\author{L.~Giannessi}
\affiliation{Elettra-Sincrotrone Trieste S.C.p.A., 34149 Basovizza, Trieste, Italy}
\author{N.~S.~Mirian}
\affiliation{Elettra-Sincrotrone Trieste S.C.p.A., 34149 Basovizza, Trieste, Italy}
\author{G.~Penco}
\affiliation{Elettra-Sincrotrone Trieste S.C.p.A., 34149 Basovizza, Trieste, Italy}
\author{G.~Perosa}
\affiliation{Elettra-Sincrotrone Trieste S.C.p.A., 34149 Basovizza, Trieste, Italy}
\affiliation{University of Trieste, Dept. Physics, 34127 Trieste, Italy}
\author{P.~Rebernik~Ribi\v{c}}
\affiliation{Elettra-Sincrotrone Trieste S.C.p.A., 34149 Basovizza, Trieste, Italy}
\author{E.~Roussel}
\affiliation{Universit\'e Lille, CNRS, UMR 8523, Physique des Lasers Atomes et Mol\'ecules, F-59000, Lille, France}
\author{I.~Setija}
\affiliation{ASML Netherlands B.V., 5504 DR Veldhoven, Netherlands}
\author{P.~Smorenburg}
\affiliation{ASML Netherlands B.V., 5504 DR Veldhoven, Netherlands}
\author{S.~Spampinati}
\affiliation{Elettra-Sincrotrone Trieste S.C.p.A., 34149 Basovizza, Trieste, Italy}
\author{C.~Spezzani}
\affiliation{Elettra-Sincrotrone Trieste S.C.p.A., 34149 Basovizza, Trieste, Italy}
\author{M.~Trov\`o}
\affiliation{Elettra-Sincrotrone Trieste S.C.p.A., 34149 Basovizza, Trieste, Italy}
\author{P.~H.~Williams}
\affiliation{ASTeC, STFC Daresbury Laboratory, Daresbury, Warrington, WA4 4AD Cheshire, United Kingdom}
\affiliation{Cockcroft Institute, Sci-Tech Daresbury, Keckwick Lane, Daresbury, Warrington, WA4 4AD, United Kingdom}
\author{A.~Wolski}
\affiliation{Cockcroft Institute, Sci-Tech Daresbury, Keckwick Lane, Daresbury, Warrington, WA4 4AD, United Kingdom}
\affiliation{Department of Physics, University of Liverpool, Liverpool, L69 7ZE, United Kingdom}
\author{S.~Di~Mitri}
\affiliation{Elettra-Sincrotrone Trieste S.C.p.A., 34149 Basovizza, Trieste, Italy}
\affiliation{University of Trieste, Dept. Physics, 34127 Trieste, Italy}

\begin{abstract} % abstract
	High-brightness electron bunches, such as those generated and accelerated in free-electron lasers (FELs), can develop small-scale structure in the longitudinal phase space. This causes variations in the slice energy spread and current profile of the bunch which then undergo amplification, in an effect known as the microbunching instability. By imposing energy spread modulations on the bunch in the low-energy section of an accelerator, using an undulator and a modulated laser pulse in the centre of a dispersive chicane, it is possible to manipulate the bunch longitudinal phase space. This allows for the control and study of the instability in unprecedented detail. We report measurements and analysis of such modulated electron bunches in the 2D spectro-temporal domain at the FERMI FEL, for three different bunch compression schemes. We also perform corresponding simulations of these experiments and show that the codes are indeed able to reproduce the measurements across a wide spectral range. This detailed experimental verification of the ability of codes to capture the essential beam dynamics of the microbunching instability will benefit the design and performance of future FELs.
\end{abstract}

\maketitle

\section{Introduction}

Laser heaters have proven to be invaluable components of high-brightness, short wavelength free-electron lasers (FELs) \cite{NIMA.528.1-2.355}, utilised in order to suppress the microbunching instability \cite{PhysRevSTAB.18.030704,NIMA.528.1-2.355,NIMA.483.1-2.268}, a collective effect that can develop due to shot noise \cite{PhysRevAccelBeams.20.054402} in the injector of such a machine, and undergo amplification due to space-charge \cite{NIMA.528.1-2.355,PhysRevSTAB.11.034401} and coherent synchrotron radiation effects \cite{PhysRevSTAB.5.064401,PhysRevSTAB.5.074401,PhysRevSTAB.12.080704}. Any small-scale structure in an electron bunch can develop and amplify during acceleration, leading to deleterious effects in the FEL process at the high-energy end of the machine. The laser heater in its nominal configuration consists of a small dispersive chicane, in the centre of which is an undulator. Co-propagating with the electron beam in the undulator is a laser pulse which imposes a correlated energy spread modulation on the beam. At the exit of the chicane, this modulation is removed due to the overlapping paths travelled in longitudinal phase space by particles with different energies. The resulting small uncorrelated energy spread increase on the bunch in the low-energy section of the accelerator, means that small-scale modulations are not able to propagate and amplify as a result of collective effects, thereby improving the FEL spectral power output. These devices are used routinely at a number of short wavelength FEL facilities to improve the quality of the light pulses produced by these machines \cite{NIMA.843.39,FEL2014.THP059,PhysRevSTAB.17.120705,FEL2017.WEP018}.

A number of schemes that impose a density or energy modulation on an electron bunch have been proposed or verified experimentally \cite{PhysRevAccelBeams.20.020706,PhysRevAccelBeams.20.050701,PhysRevAccelBeams.22.060701,PhysRevLett.111.134802,PhysRevLett.108.144801,NatPhys.4.390,PhysPlasmas.25.043111,PhysRevLett.122.044801,PhysRevLett.111.034803,PhysRevLett.116.184801,PhysRevSTAB.16.100701}, either by passing the bunch through a wakefield-generating structure, taking advantage of longitudinal space-charge oscillations, or by manipulating the electron bunch directly with a laser pulse, either at the point of generation or further down the accelerator. Methods of imposing \si{\tera\hertz}-scale modulations on an electron bunch could have wide-ranging applications \cite{NatPhoton.1.97}, from manipulating and accelerating particle beams \cite{NatComms.6.8486,JPhysB.51.20.204001,NatPhoton.12.336} to scientific and industrial imaging applications \cite{ApplOpt.49.19.E48,NatPhoton.8.605}. 

Recent experiments at the FERMI FEL \cite{NatPhoton.6.699} have investigated the possibility of using a laser pulse with a non-uniform longitudinal intensity profile to impose energy spread modulations on the bunch in the laser heater, thereby seeding the microbunching at a known single frequency \cite{PhysRevLett.115.214801,IPAC2017.THYA1}. This is achieved through chirped-pulse beating, in which the laser pulse is chirped and split in an interferometer, then recombined with a variable delay on one of the interferometer arms \cite{JOptSocAmB.13.12.2783}. The beating frequency of this chirped pulse is then proportional to the delay between the two pulses. This study builds on previous work done on premodulated beams in storage rings \cite{NewJPhys.16.063027}. By applying this technique to the laser pulse used for the laser heater in its nominal configuration, the beating wavelength of this modified pulse can be increased by orders of magnitude, thus generating a laser pulse with an effective wavelength of a similar order to the length of the electron bunch. The imposition of such a laser pulse onto the electron bunch initially causes an energy spread modulation, which is then mixed into the longitudinal phase space such that there exists a modulation in both energy and longitudinal density at the exit of the laser heater chicane. This effect is then further enhanced through magnetic bunch length compression. 

In this paper we study the effect of imposing such a modulated laser pulse on the electron beam in the FERMI laser heater for three bunch compression scenarios, and for a range of laser pulse modulation frequencies and laser pulse energies. By analysis of the full longitudinal phase space, rather than only the current density, a more detailed picture of the microbunching instability can be obtained, as the evolution of the microbunching instability through the oscillation between energy and density modulations can be investigated in full \cite{SciRep.10.5059}. 

An additional long-standing problem concerns the reliability of simulation codes in terms of their ability to reproduce the effect of the microbunching instability, being a result of the collective interaction of a large number of particles which occurs on a small length scale \cite{NIMA.483.1-2.268,PhysRevAccelBeams.20.054402}. Full start-to-end simulations of the experiments reported in this paper have been conducted in order to assess the capability of existing codes to model this longitudinal phase-space manipulation, such that the measured bunching at the end of the accelerator can be reflected in simulations. Accurate and reliable simulations are essential for the design of systems and future experiments that require control over the bunch profile. Our work comprises a systematic, accurate and quantitative characterization of the microbunching instability in codes and in measurements across a wide range of machine configurations. By benchmarking the codes against experimental measurements, we have demonstrated the feasibility of simulating microbunching over a range of conditions. 

The paper is laid out as follows. In Section\,\ref{sec:microbunching_instability} a brief review of the theory of the microbunching instability is given. The accelerator lattice and bunch parameters are described in Section\,\ref{sec:machineconfig}, and the chirped-pulse beating setup is outlined in Section\,\ref{sec:cpb}. Section\,\ref{sec:simulations} gives an overview of the codes used for simulating the experiment. Some examples of longitudinal phase space images at the linac exit for the various lattice configurations are shown in Section\,\ref{sec:measurements}, and in Section\,\ref{sec:benchmarking} the microbunching parameters for each of these bunch compression schemes are analysed, and compared with results from simulations.

\section{Microbunching Instability}\label{sec:microbunching_instability}

%Small-scale structure in the electron bunches of FELs can arise due to shot noise in the low energy-regime of the accelerator, the interplay of the rise-time and non-uniformity of the photoinjector laser pulse, and the response time of the photocathode \cite{PhysRevAccelBeams.23.024401}. This structure can then undergo amplification due to impedance effects such as longitudinal space-charge (LSC) forces, \cite{PhysRevSTAB.7.074401,NIMA.528.1-2.355} and coherent synchrotron radiation (CSR) \cite{PhysRevSTAB.5.074401,PhysRevSTAB.12.080704,PhysRevSTAB.5.064401} in dispersive regions. This instability is of critical importance for high-brightness electron sources \cite{PhysRevSTAB.13.020703,PhysRevSTAB.5.074401,NIMA.483.1-2.268,PhysRevSTAB.11.030703,PhysRevAccelBeams.20.120701}, as it can both disable diagnostics devices, and cause a degradation in the beam quality, which then has an adverse effect on the application of the beam, such as in an FEL.
 
Previous experimental studies of the microbunching instability have analysed the structure along the longitudinal plane of the bunch \cite{PhysRevSTAB.18.030704}; here, we use the procedure introduced in Ref.\,\cite{SciRep.10.5059} to analyse the bunching in both time and energy. An example of the significance of the second of these parameters is in a free-electron laser: if there are discrete energy bands in an electron bunch upon its entrance to the undulator, this may result in the generation of multicolour photon pulses. This is analogous to the process that occurs in the first modulator and chicane of the echo-enabled harmonic generation scheme \cite{PhysRevLett.102.074801}. As a result of longitudinal space charge (LSC) forces, there is a plasma oscillation between energy and density modulations that develops along the machine \cite{NIMA.393.1-3.376,PhysRevLett.106.184801}, and so the final longitudinal phase space orientation of the microbunches will depend on the plasma oscillation phase at this point. Analysis of the full longitudinal phase space allows the plasma oscillation phase at the measurement point to be determined.

Here, we briefly review the theory of microbunching due to LSC and coherent synchrotron radiation (CSR). Since the space-charge field experienced by an electron in the bunch is dependent on neighbouring electrons, small deviations in the density profile can cause a change in energy for a given electron. In the vicinity of a peak in the longitudinal charge density, space-charge forces can accelerate particles ahead of the peak and decelerate those behind.  Over time, this changes a modulation in the charge density to a modulation in the energy as a function of longitudinal position in the bunch.  Small variations in particle velocity (in low or medium energy bunches) then change the energy modulation back into a density modulation.

Due to the mutual repulsion of electrons in a high-density region, and the subsequent variation in electric field that causes acceleration and deceleration of particles, a longitudinal space-charge oscillation between energy and density modulations takes place, which depends on the LSC impedance $Z_{LSC}$ \cite{PhysRevSTAB.11.034401,PhysRevSTAB.12.100702}. In the case of a relativistic beam in a drift space, this oscillation has a characteristic frequency given by \cite{PhysRevSTAB.7.074401}:

\begin{equation}
\omega_{LSC} = c \left[ \frac{I_0}{\gamma^3 I_A} k_0 \frac{4\pi |Z_{LSC}(k_0)|}{Z_0} \right]^{1/2} \lesssim \frac{2c}{r_b} \left(\frac{I_0}{\gamma^3 I_A}\right)^{1/2},
\end{equation}

\noindent with an initial modulation wavenumber $k_0$ and $I_A \approx 17$\,\si{\kilo\ampere} the Alfv\'en current. The initial peak current $I_0 = n e c$ for $n$ electrons per unit length, $e$ the electron charge and $c$ the speed of light. Since this oscillation frequency is inversely proportional to the beam energy, the period of the oscillation between energy and density modulations can become large (on the order of $10$s or $100$s of metres) for ultrarelativistic beams. Nevertheless, due to the long accelerating sections required to drive a short wavelength FEL, this oscillation cannot be neglected when considering the evolution of the microbunching instability in such machines. 

CSR emitted by a short bunch in a magnetic compressor can cause self-interactions between the electrons in the bunch, causing an increase in both the projected emittance and correlated energy spread \cite{NewJPhys.20.073035}. In the case of a bunch with density perturbations in the longitudinal phase space, there can be strong CSR emission at the wavelength of the density modulation, inducing a modulation in energy. The longitudinal dispersion $R_{56}$ of the bunch compressor -- which relates the change in longitudinal position of a particle (relative to a reference particle) to its energy -- can then convert these energy modulations into density modulations, thereby enhancing the microbunching in longitudinal density. 

The CSR impedance increases monotonically (but weakly) with increasing $k$: i.e. the impedance (and hence the energy modulation resulting from a given density modulation) is larger at larger $k$, i.e. at smaller wavelength. In a bunch compression chicane, the longitudinal density modulation is no longer fixed, as the longitudinal motion in a dipole is coupled to the horizontal motion, and the modulation wavelength reduces with the bunch length. 

Taking all of these effects into account, the microbunching in the longitudinal phase space of a particle bunch can be described using the bunching factor $b$ \cite{PhysRevSTAB.5.074401}. When analysing microbunching in the longitudinal plane only, this bunching factor is described by the Fourier transform of the current density of the bunch, which is influenced by the impedance functions mentioned above. In order to extend this analysis to two dimensions, we simply take the two-dimensional Fourier transform of the longitudinal phase space density, $\rho_{t,E}$:

\begin{equation}
\label{eq:bf}
b(k, m) = \frac{1}{N} \int \int \rho_{t,E} e^{-i\left(k t + m E \right)} dt dE,
\end{equation}

\noindent where $N$ is the number of particles, and $k$ and $m$ describe, respectively, the frequency modulation in the temporal and energy planes \cite{SciRep.10.5059}.

\section{Accelerator Parameters} \label{sec:machineconfig}

\begin{figure}[bth!]
	\begin{center}
		\includegraphics[width=8.6cm]{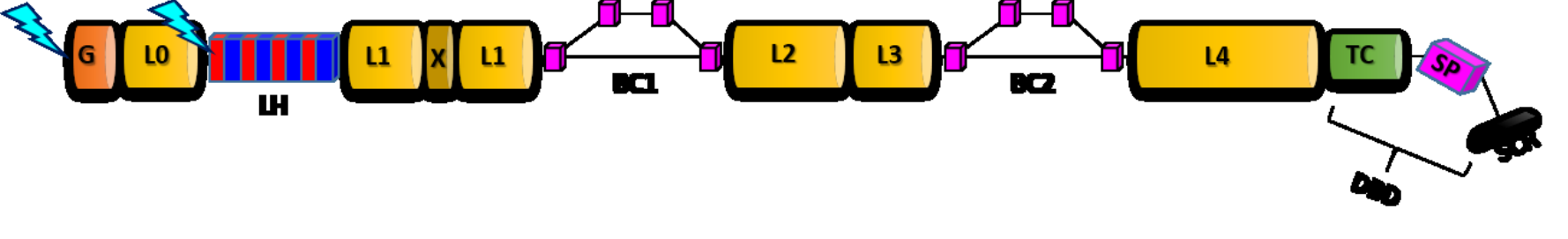}
		\caption{Schematic of the FERMI linac (not to scale). The beam is produced and accelerated initially in the gun (G), and is subsequently accelerated in linacs L0 \mbox{--} 4. The laser heater (LH), used for manipulating the electron bunch longitudinal phase space, is located between L0 and L1. The two variable bunch compressors are labelled as BC1 and BC2. At the exit of L4, the beam is streaked via the transverse deflecting cavity (TC) and observed in the diagnostics beam dump (DBD) line after passing through a spectrometer dipole (SP) and being imaged on a screen (SCR).} \label{fig:fermi_linac}
	\end{center}
\end{figure}

A schematic of the FERMI linac is shown in Fig.\,\ref{fig:fermi_linac}. Electrons are produced in a high-brightness electron gun, and accelerated in Linac 0 (L0) to around $100$\,\si{\mega\electronvolt}. The laser heater consists of a small dispersive chicane, in the centre of which is a short planar undulator. Within this undulator, an infrared laser beam is superimposed temporally and spatially onto the electron bunch, thus producing an energy spread modulation on a scale proportional to the laser wavelength. This modulation is then removed as the bunch exits the second half of the laser heater chicane. The laser heater parameters are given in Table\,\ref{table:lh_parameter_table} -- for more details on the system, see Ref.\,\cite{PhysRevSTAB.17.120705}. 

\begin{table}[bth!]
	\centering
	\caption{Laser heater system parameters} % title of Table
	\label{table:lh_parameter_table}
	\begin{tabular}{ll}
		\hline
		\multicolumn{2}{c}{\textsc{BEAM TRANSPORT}} \\
		\hline
		Chicane magnet bend angle 	 & $3.5$\,\si{\degree} \\
		Transverse offset in chicane & $30$\,\si{\milli\metre} \\
		Dispersion	 				 & $30$\,\si{\centi\metre} \\
		Beam energy  				 & $96$\,\si{\mega\electronvolt} \\
		Emittance 					 & $0.35$\,\si{\milli\metre-\milli\radian} \\
		Transverse beam size 		 & $70$\,\si{\micro\metre} \\
		\hline
		\multicolumn{2}{c}{\textsc{UNDULATOR}} \\
		\hline
		Period 						 & $40$\,\si{\milli\metre} \\
		Number of full periods 		 & $12$ \\
		Undulator parameter $K$ 	 & $0.88$ \\
		\hline
		\multicolumn{2}{c}{\textsc{LASER}} \\
		\hline
		Wavelength 					 & $783$\,\si{\nano\metre} \\
		Spot size 					 & $120$\,\si{\micro\metre} \\
		Pulse length (FWHM)			 & $16.5$\,\si{\pico\second} \\
		Pulse energy (maxmimum)		 & $20$\,\si{\micro\joule} \\
		Spectral bandwidth 			 & $5$\,\si{\nano\metre} \\
		Linear chirp coefficient	 & $-1.5 \times 10^{23}$\,\si{\second}$^{-2}$ \\
        Delay between pulses & $4 \hbox{--} 30$\,\si{\pico\second} \\
		\hline
	\end{tabular}
\end{table}  

At the exit of the laser heater section, the bunch is accelerated in Linac 1 (L1) to an energy of around $300$\,\si{\mega\electronvolt} -- this accelerating section also includes an X-band cavity to linearise the longitudinal phase space, and therefore the bunch compression process. The first variable bunch compressor, BC1, is located at the exit of this linac, after which point the bunch is further accelerated in the remaining accelerating sections, Linacs 2, 3, and 4. A second variable bunch compressor, BC2, is located between L3 and L4. After L4, there is a full beam diagnostics suite, including a vertically deflecting RF cavity and a dipole magnet for longitudinal phase space measurements \cite{IEEETransNuclSci.62.1.210}. For the purposes of our experiment, L4 was switched off, meaning that the final beam energy at the diagnostic point was around $715$\,\si{\mega\electronvolt} for one of the compression schemes, and around $780$\,\si{\mega\electronvolt} for the other two. The temporal and energy resolutions provided at the diagnostics station were around $10$\,\si{\femto\second} and $100$\,\si{\kilo\electronvolt}, respectively.

Measurements of the bunch longitudinal phase space were made at the diagnostics beam dump (DBD) screen. Three different machine configurations were employed for this study, corresponding to different combinations of bunch compressors used to compress the beam: BC1 only, BC2 only, and a combination of BC1 with BC2 -- the lattice and beam parameters are summarised in Table\,\ref{table:fermi_parameters}. For each of these three lattice configurations, the longitudinal phase space was measured for a number of settings of the laser heater, including variations in power and intensity modulation wavelength.%For the BC1 only case, the bending angles of the chicane dipoles were set to $105$\,\si{\milli\radian}; for BC2 only, the bending angles were set to $90$\,\si{\milli\radian}; for BC1 and BC2 together, the bending angle of the first bunch compressor was set to $105$\,\si{\milli\radian} and that of the second was set to $28.5$\,\si{\milli\radian}. For each of these three lattice configurations, the longitudinal phase space was measured for a number of settings of the laser heater power.%was varied between $17.1$ and $35.2$\,\si{\milli\radian}. For each of these three lattice configurations, the longitudinal phase space was measured for a number of settings of the laser heater power.

\begin{table}
\caption{Main beam parameters of the FERMI accelerator at the end of Linac 4 for the three compression schemes.}
\centering
\label{table:fermi_parameters}
\begin{tabular}{|c|c|c|c|}
	\hline
	\textbf{Bunch parameters}     & \textbf{BC1 only} 		 		 & \textbf{BC2 only} 		 	    & \textbf{BC1+BC2} \\
	\hline%\\ [0.2ex]	 
	Bunch charge (\si{\pico\coulomb})	   		      & $100$ 		 & $100$ 	    & $100$  	   \\
	Beam energy  (\si{\mega\electronvolt})	       	  & $775$		 & $715$	    & $780$  \\
	%RMS bunch length  (\si{\femto\second})	       	  & $42.0$		 & $54.2$	    & $34.7$  \\
	Chicane bending angle (\si{\milli\radian})         & $105$ 		 & $90$  	    & $105$ + $28.5$ \\
	Relative energy spread $dE/E$ ($\%$) & $0.5$					 & $0.4$					    & $0.05$ \\
	Longitudinal dispersion $R_{56}$ (\si{\milli\metre})  & $-20.9$   	 & $-10.6$   	& $-22.5, -3.5$ \\
	Peak current (\si{\ampere})	& $620$ 		 	 & $540$		 	    & $650$ \\
	%Projected $\epsilon_{N,x}$ (\si{\micro\metre\radian})  & $2.68$ & $0.93$ & $2.07$ \\
	%Projected $\epsilon_{N,y}$ (\si{\micro\metre\radian})  & $0.74$ & $0.86$ & $4.44$ \\
	Energy chirp ($\%/$\si{\milli\metre})	& $-8.7$  & $-5.6$ & $-1.0$ \\
	\hline
\end{tabular}
\end{table}
\section{Chirped-Pulse Beating}\label{sec:cpb}

Modulation of the laser heater pulse in FERMI is achieved through chirped-pulse beating \cite{JOptSocAmB.13.12.2783}. The initial pulse (Gaussian in both the transverse and longitudinal dimensions) is stretched temporally (or \lq chirped\rq), then split in a Michelson interferometer, one arm of which has a variable length. The two laser pulses are recombined, and they overlap in the temporal domain. By varying the length of the interferometer arm, a delay between the two pulses can be created, giving rise to a laser pulse with a beat frequency that is directly related to the delay $\tau$. The intensity profile of such a laser pulse is given by \cite{NatPhys.4.390,PhysRevSTAB.13.090703}:

\begin{equation}\label{eq:lhbeating}
\begin{split}
I_{tot}(t, \tau) = {E_0}^2\left(\frac{\Delta t_0}{\Delta t_1}\right) \left[ e^{\left(-2a_1(\Delta t_1)\left(t + \tau /2 \right)^2\right)} + e^{\left(-2a_1(\Delta t_1)\left(t - \tau /2 \right)^2\right)} \right. \\
\left. + \left( e^{\left(-2a_1(\Delta t_1)\left(t + \tau /2 \right)^2\right)}\cos\left(\omega_0 \tau + 2 b_1(\Delta t_0 \Delta t_1) t \tau \right) \right) \right],
\end{split}
\end{equation}

\noindent with $E_0$ the field strength of the initial pulse, $\Delta t_0$ the Gaussian half-width of the initial pulse, $\Delta t_1$ the stretched pulse half-width, $\omega_0$ the centre of the optical pulse spectrum, and

\begin{subequations}
	\begin{equation}
	a_{1}(\Delta t_1) = \frac{2 \ln (2)}{\Delta t_1 ^2},
	\end{equation}
	\begin{equation}
	b_{1}(\Delta t_0, \Delta t_1) = \frac{2 \ln (2)}{\Delta t_0 \Delta t_1}.
	\end{equation}
\end{subequations}

\begin{figure}[bth!]
	\begin{center}
		\centering
		\subfloat[]{  
			\includegraphics[width=8.6cm]{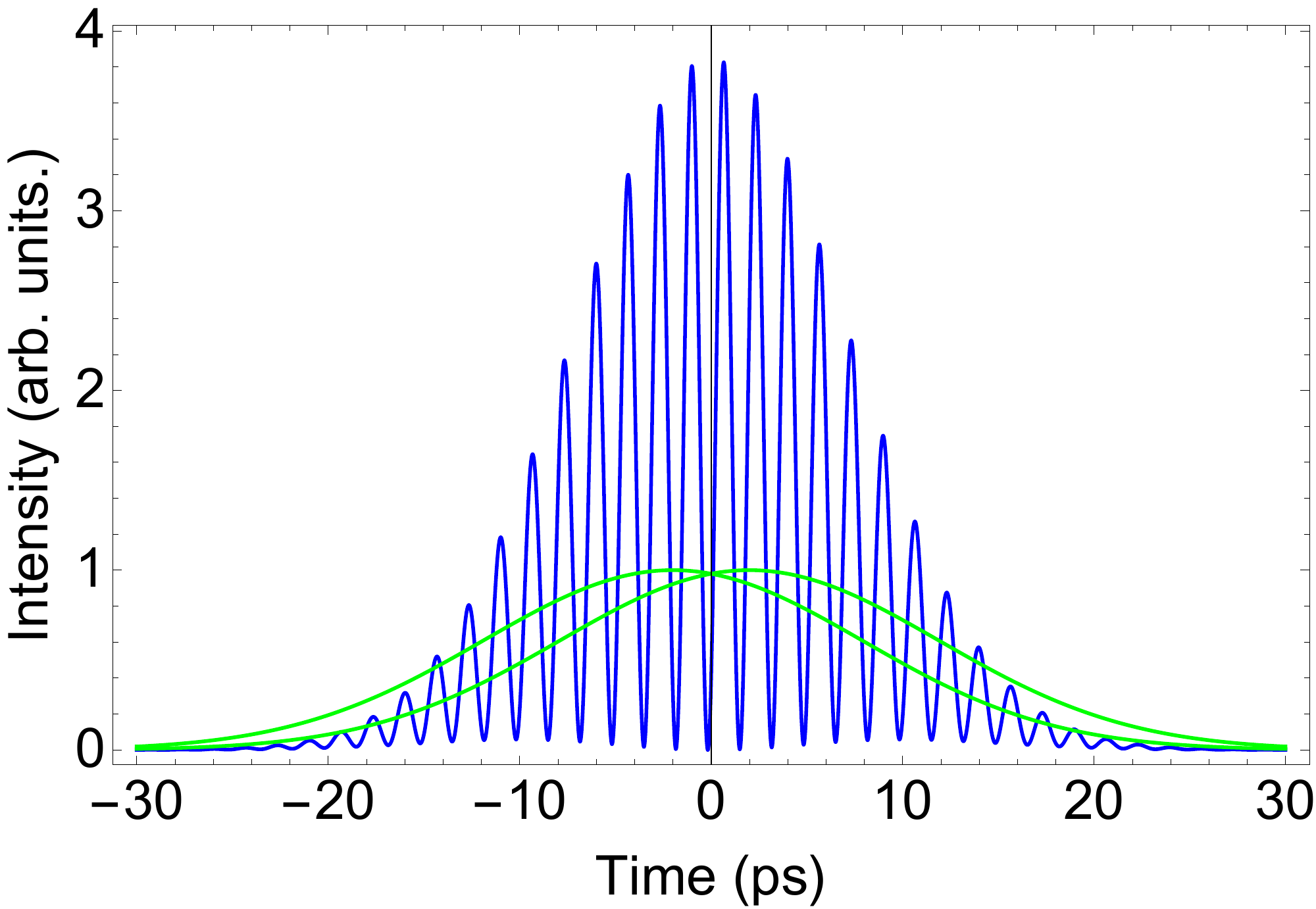} 
			\label{fig:beatplots_4}}
		\vfill
		\subfloat[]{  
			\includegraphics[width=8.6cm]{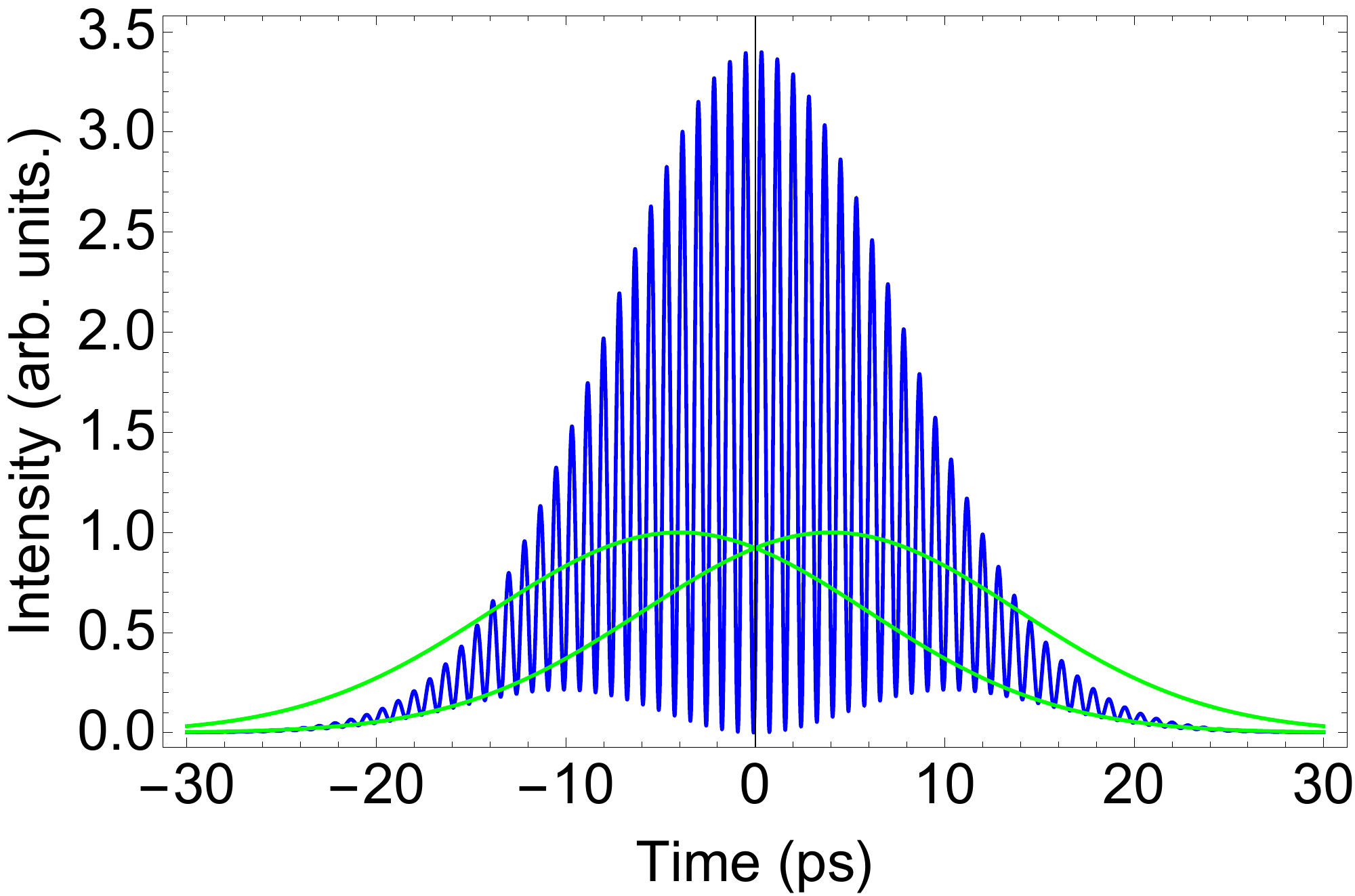} 
			\label{fig:beatplots_8}}
		\vfill
		\subfloat[]{  
			\includegraphics[width=8.6cm]{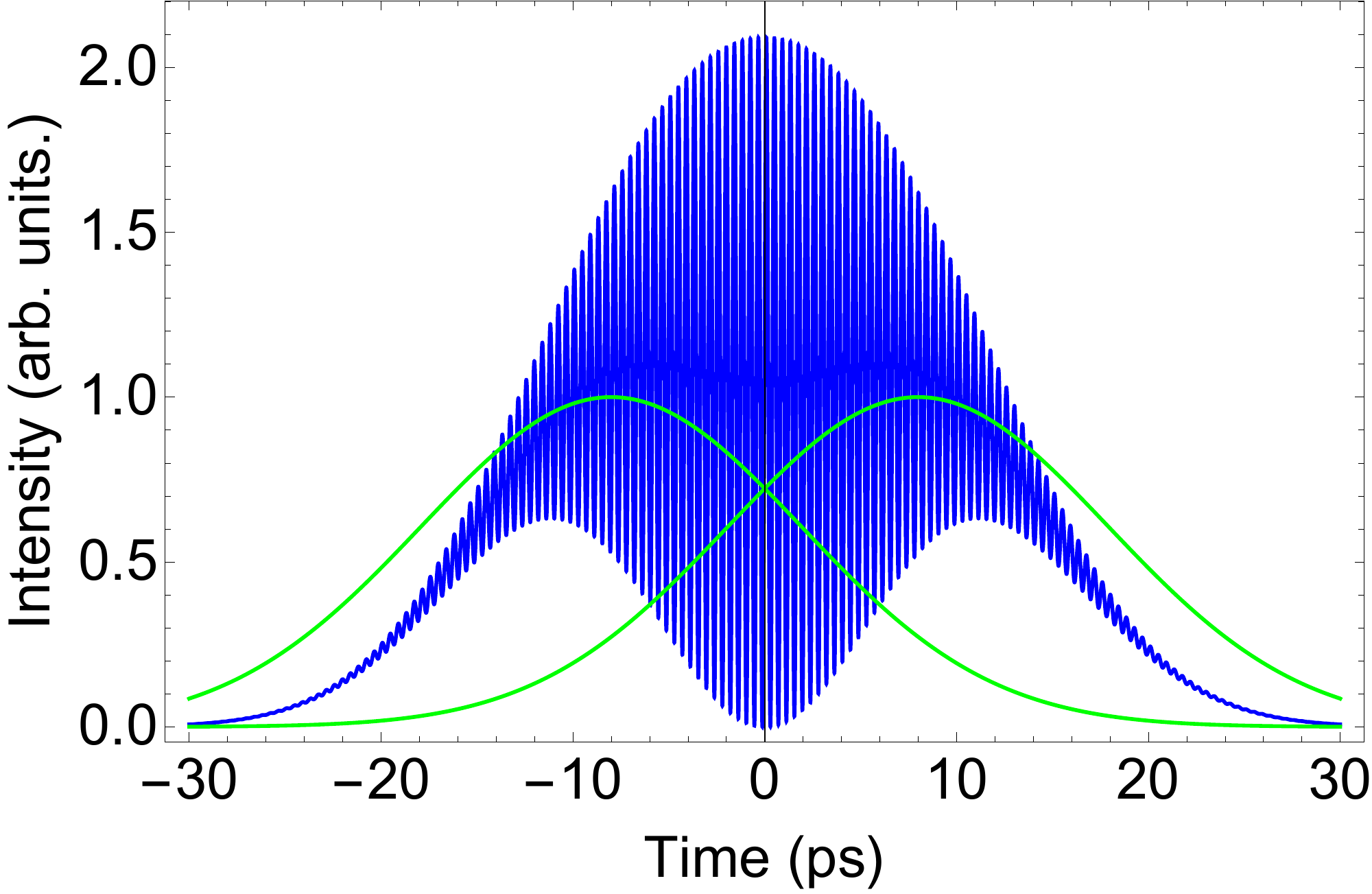} 
			\label{fig:beatplots_16}}
		\caption{Calculated intensity profiles (blue) of modulated laser pulses (from Eq.\,\ref{eq:lhbeating}) for delays between pulses of: above: $4$\,\si{\pico\second}; middle: $8$\si{\pico\second}; bottom: $16$\si{\pico\second}. The intensity profiles for the two separated laser pulses before recombination are shown in green.} \label{fig:chirped_pulse_beating}
	\end{center}
\end{figure}

\iffalse
\onecolumngrid

\begin{figure}[bth!]
	\begin{center}
		\includegraphics[width=17.2cm]{beatplots.pdf}
		\caption{Calculated intensity profiles of modulated laser pulses (from Eq.\,\ref{eq:lhbeating}) for delays between pulses of: Left: $4$\,\si{\pico\second}; middle: $8$\si{\pico\second}; right: $16$\si{\pico\second} (Blue). The intensity profiles for the two separated laser pulses before recombination are shown in Green and Orange.} \label{fig:chirped_pulse_beating}
	\end{center}
\end{figure}

\twocolumngrid
\fi
\noindent For a frequency chirp rate of $\mu$ (i.e. the rate of change of frequency with time), the beat frequency of the modulated laser is given by $f(\tau, \mu)\approx \mu \tau / 2 \pi$ \cite{JOptSocAmB.13.12.2783}. Using the parameters given in Table\,\ref{table:lh_parameter_table}, we calculate the beating frequency of the laser as $\alpha \cdot \tau$, where $\alpha = \frac{-\Delta \lambda c}{\lambda^2 \Delta \sigma}$ is the linear chirp coefficient for a laser pulse with bandwidth $\Delta \lambda$ and $\Delta \sigma$ the difference in pulse length between the stretched and compressed pulses. The final modulation $\lambda_{f}$ on the bunch after compression is then equivalent to the initial modulation $\lambda_{i}$ divided by the compression factor $C$, which in this case is approximately in the range $20 \hbox{--} 25$ for the various compression schemes.

By applying the chirped-pulse beating technique, we can obtain a range of longitudinal laser intensity profiles. Some examples are shown in Fig.\,\ref{fig:chirped_pulse_beating}. The flexibility of laser intensity modulations provided by this technique can lead to the generation of a range of customisable longitudinal electron beam profiles.

\begin{figure}[bth!]
	\begin{center}
		\includegraphics[width=8.6cm]{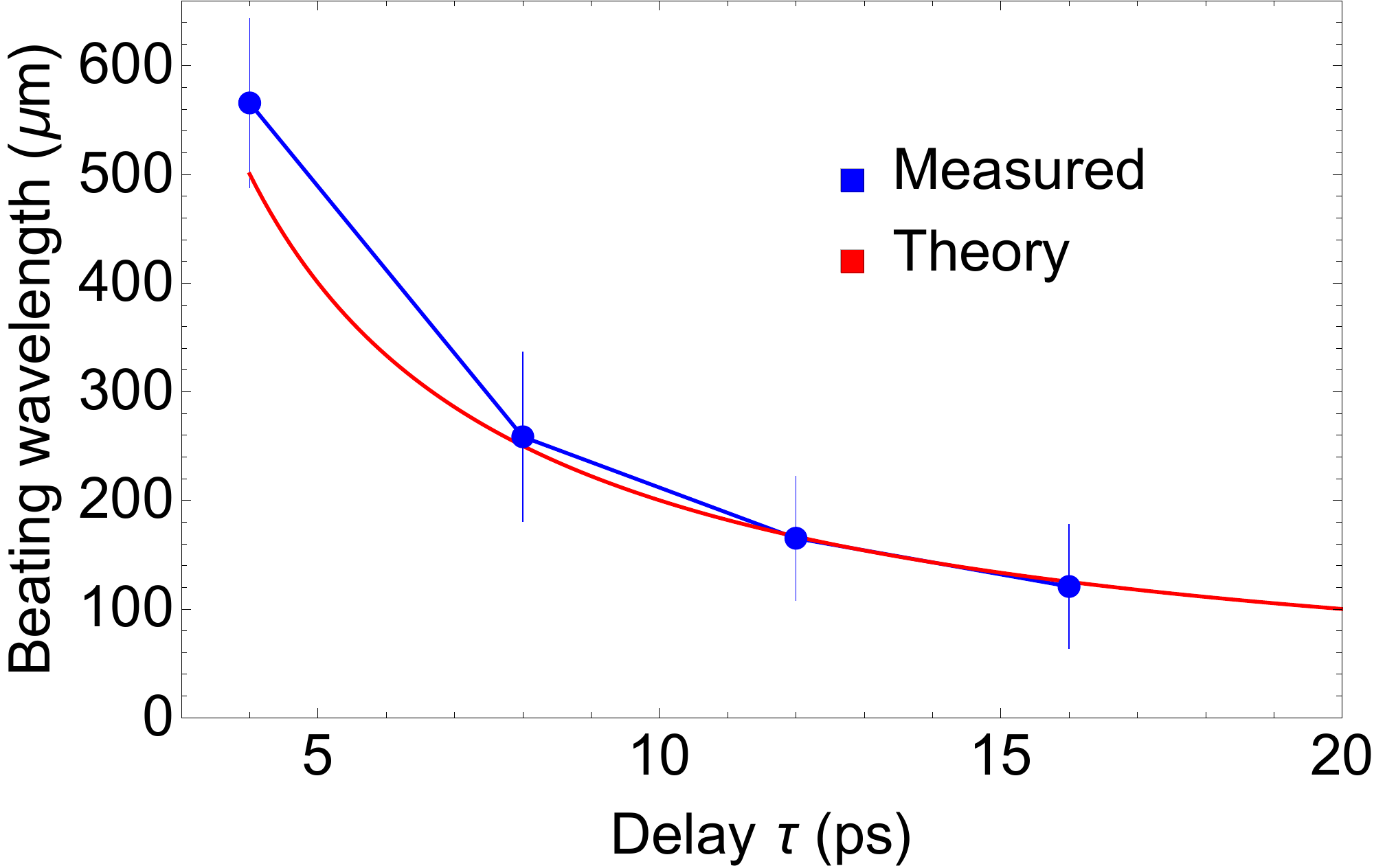}
		\caption{Measured (blue) and predicted (red) modulation period on the electron bunch as a function of interferometer delay $\tau$, with an uncompressed bunch.} \label{fig:delay_vs_wavelength}
	\end{center}
\end{figure}

In order to cross-check the measured electron bunch modulation period with the beating wavelength of the laser pulse, measurements of the variation in energy spread along the bunch were taken at the low-energy RF deflecting cavity, located at the exit of BC1 (see Fig.\,\ref{fig:fermi_linac}). In this case, the bunch was uncompressed. The modulation period along the length of the bunch was extracted for a range of values of the laser beating delay $\tau$. Fig.\,\ref{fig:delay_vs_wavelength} shows the variation of delay between pulses and the corresponding modulation period on these bunches, along with values for the beating wavelength of the laser pulse from the theory. It can be seen that the measurements agree well with the predictions. With a variation of the delay $\tau$ between $4$\,\si{\pico\second} and $30$\,\si{\pico\second} -- corresponding to initial modulations imposed on the bunch in the range $0.6 \hbox{--} 4.5$\,\si{\tera\hertz} -- we are able to probe a wide range of modulated longitudinal phase spaces.

\section{Simulation Tools}\label{sec:simulations}

Simulations of the FERMI injector (up to the exit of Linac 0, see Fig.\,\ref{fig:fermi_linac}) have been produced using the General Particle Tracer (\textsc{GPT}) code \cite{GPT}. In order to match accurately the simulation to experimental conditions, the measured transverse and longitudinal profiles of the photoinjector laser were used as input parameters to the simulation, along with geometric wakefields from the injector linac. Full 3D space-charge effects were also included. The injector linac phase was optimised for minimal energy spread -- as is done in the routine procedure of linac tuning -- and good agreement was found between the simulated and experimentally measured bunch properties at the exit of the injector.

From this injector simulation, the bunch is smoothed using a Poisson distribution, in order to simulate the shot noise arising in the low-energy section of the machine. The bunch is then tracked using the \textsc{Elegant} code \cite{APS-LS-287} through the rest of the machine up to the DBD screen. Effects of linac wakefields \cite{PhysRevAccelBeams.22.014401,NIMA.558.1.58} and the laser heater, and 1D CSR \cite{PhysRevSTAB.4.070701} and LSC models \cite{PhysRevSTAB.7.074401} are included in the tracking. The code computes the impedances of these collective effects based on a binned histogram of the longitudinal density distribution. The effect of the transverse deflector and propagation of the beam to the diagnostics line is also included, providing a complete simulation of the measurement.

\iffalse
It should be mentioned that the number of bins used for the density histogram in the CSR and longitudinal space-charge (LSC) models of \textsc{Elegant}, in addition to the smoothing applied on the bunch, can have an impact on the final results \cite{PhysRevSTAB.4.070701}. Following a convergence study (based on a procedure outlined in \cite{OAG-TN-2005-27}), by varying the number of LSC and CSR bins between $100$ and $5000$, for a range of macroparticle numbers between $10^5$ and $2 \times 10^7$ macroparticles, we observe a variation in the bunching factor observed on the final screen.
\fi

By comparing the measured and simulated bunching factors, it is possible to determine an optimal setting for the number of bins (and the number of macroparticles) to be used in simulations of the microbunching instability. In our case, since the ratio between the initial density modulation wavelength and the bunch length in the laser heater undulator is known, it is possible for the simulation code to find a balance between under- and over-estimating the effect of CSR and LSC \cite{PhysRevSTAB.11.030701}. The user can also apply a high-pass filter in order to control the effect of numerical noise in the simulation.

The \textsc{Elegant} code also provides the functionality to simulate the interaction between an electron and a user-defined laser pulse in an undulator. By tracking a bunch through the laser heater undulator using pulses generated based on the parameters given in Table \ref{table:lh_parameter_table} and Eq.\,\ref{eq:lhbeating}, it is possible to reproduce the measured effect of a temporally modulated laser heater pulse on the electron beam.

\section{Induced Modulation in Different Compression Schemes}\label{sec:measurements}

%Measurements of the longitudinal phase space were done using a vertical RF deflecting cavity and a bending magnet. The beam is streaked using a transverse mode RF field with the phase set such that the beam experiences a vertical streaking, therefore imposing a time-energy correlation in this direction. When coupled with a horizontal bend, creating a correlation between energy and horizontal position, the transverse distribution of the bunch (imaged with a screen) represents the longitudinal phase space of the bunch \cite{IEEETransNuclSci.62.1.210}. For every lattice configuration, the beam optics were matched on the entrance to the deflector such that the resolution in both planes was optimised. The resolution varied slightly between different lattice configurations, but the temporal and energy resolution remained around $10$\,\si{\femto\second} and $80 \hbox{--} 100$\,\si{\kilo\electronvolt}, respectively, throughout the experiment.

The electron beam longitudinal phase space was measured for a number of machine configurations: in addition to varying the bunch compression process using combinations of the two variable bunch compressors, the microbunching was seeded in the laser heater for a range of initial modulation wavelengths and laser pulse energies. The timing between the electron beam and the laser in the laser heater chicane can also be varied, allowing for timing scans that provided the largest amplification of the microbunches. \iffalse Since the microbunching amplification undergone by the bunch is a function of modulation wavelength, the maximum bunching factor for a certain laser pulse energy in the laser heater undulator is related to the largest microbunching gain. \fi In this section, we present some sample measurements of the longitudinal phase space for three bunch compression schemes and their associated Fourier transforms.

\subsection{Double Compression}\label{subsec:double_compression}

\begin{figure*}[bth!]
	\begin{center}
		\centering
		\subfloat[]{  
			\includegraphics[width=4.3cm]{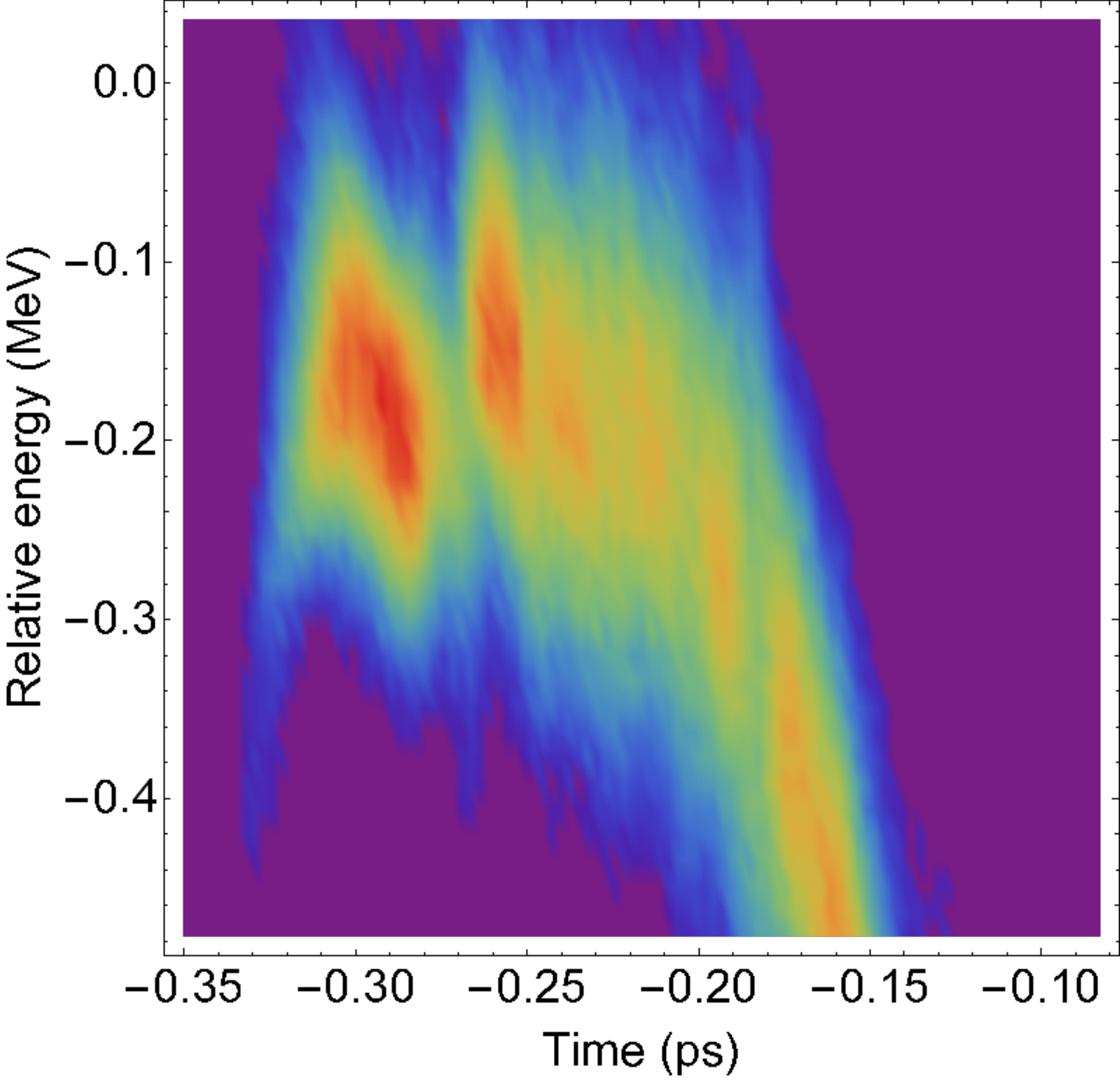} 
			\label{fig:lpsbc1bc28500112ps}}
		\subfloat[]{  
			\includegraphics[width=4.3cm]{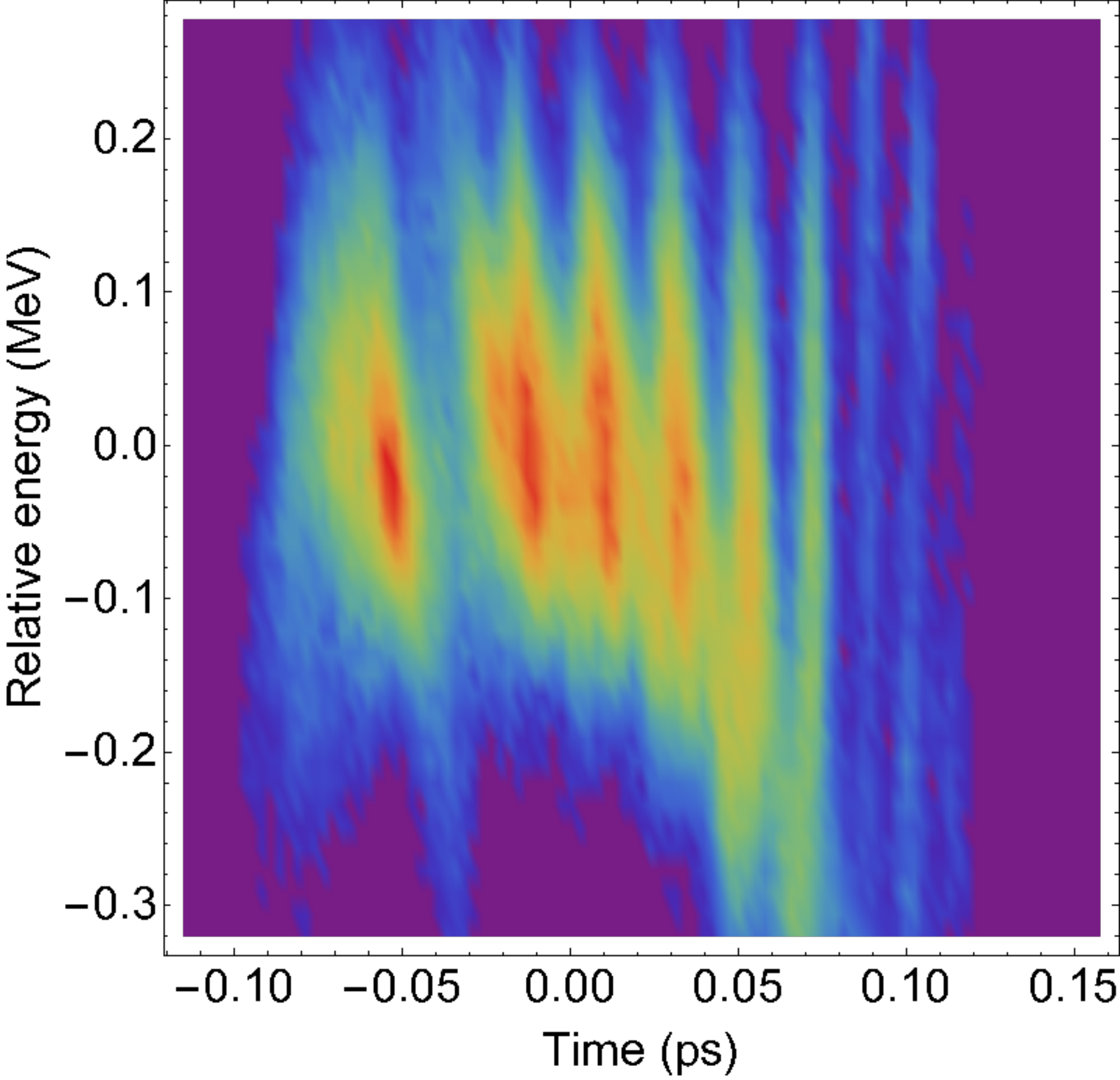} 
			\label{fig:lpsbc1bc28505112ps}}
		\subfloat[]{  
			\includegraphics[width=4.3cm]{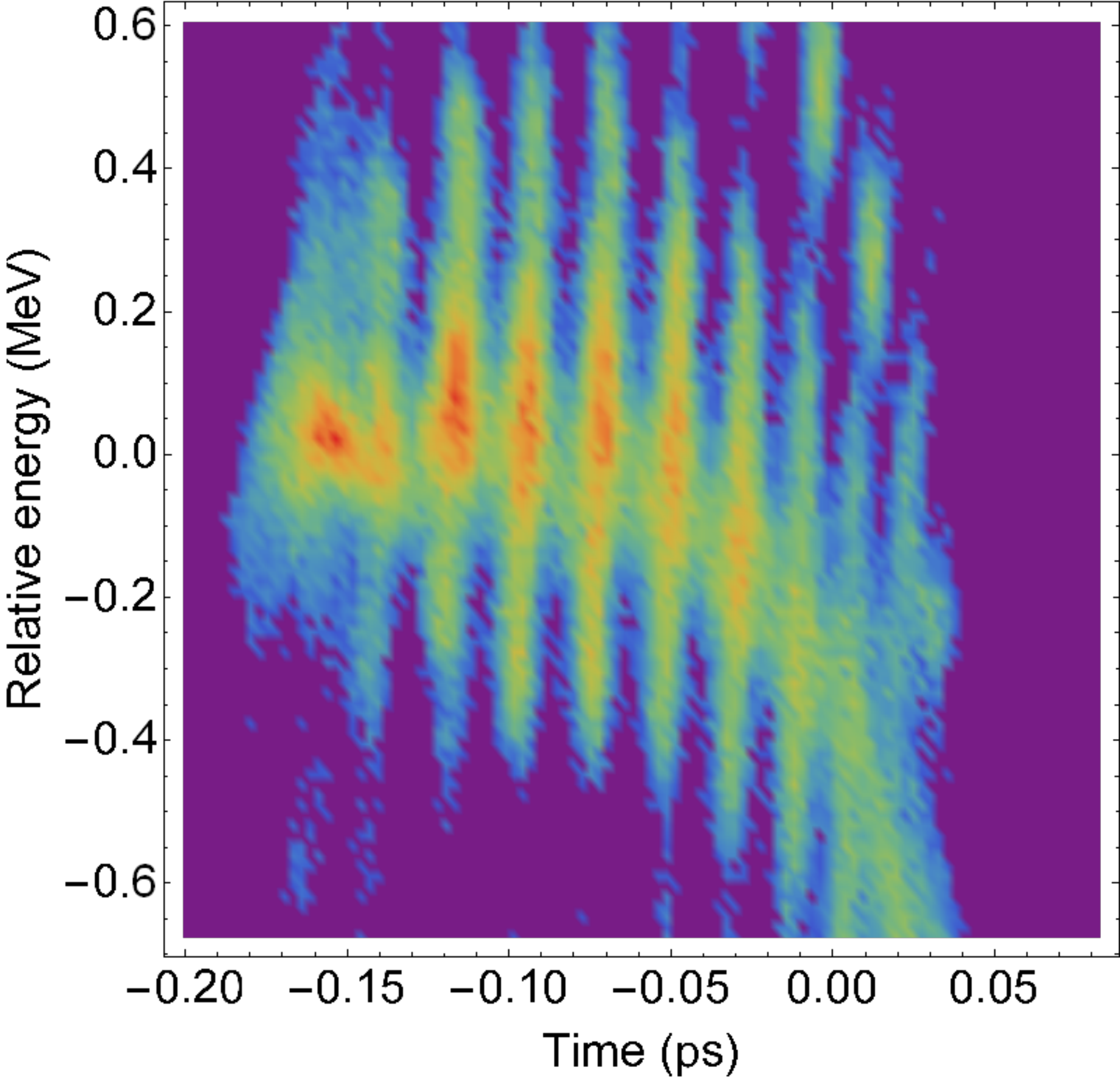} 
			\label{fig:lpsbc1bc28501012ps}}
		\subfloat[]{  
			\includegraphics[width=4.3cm]{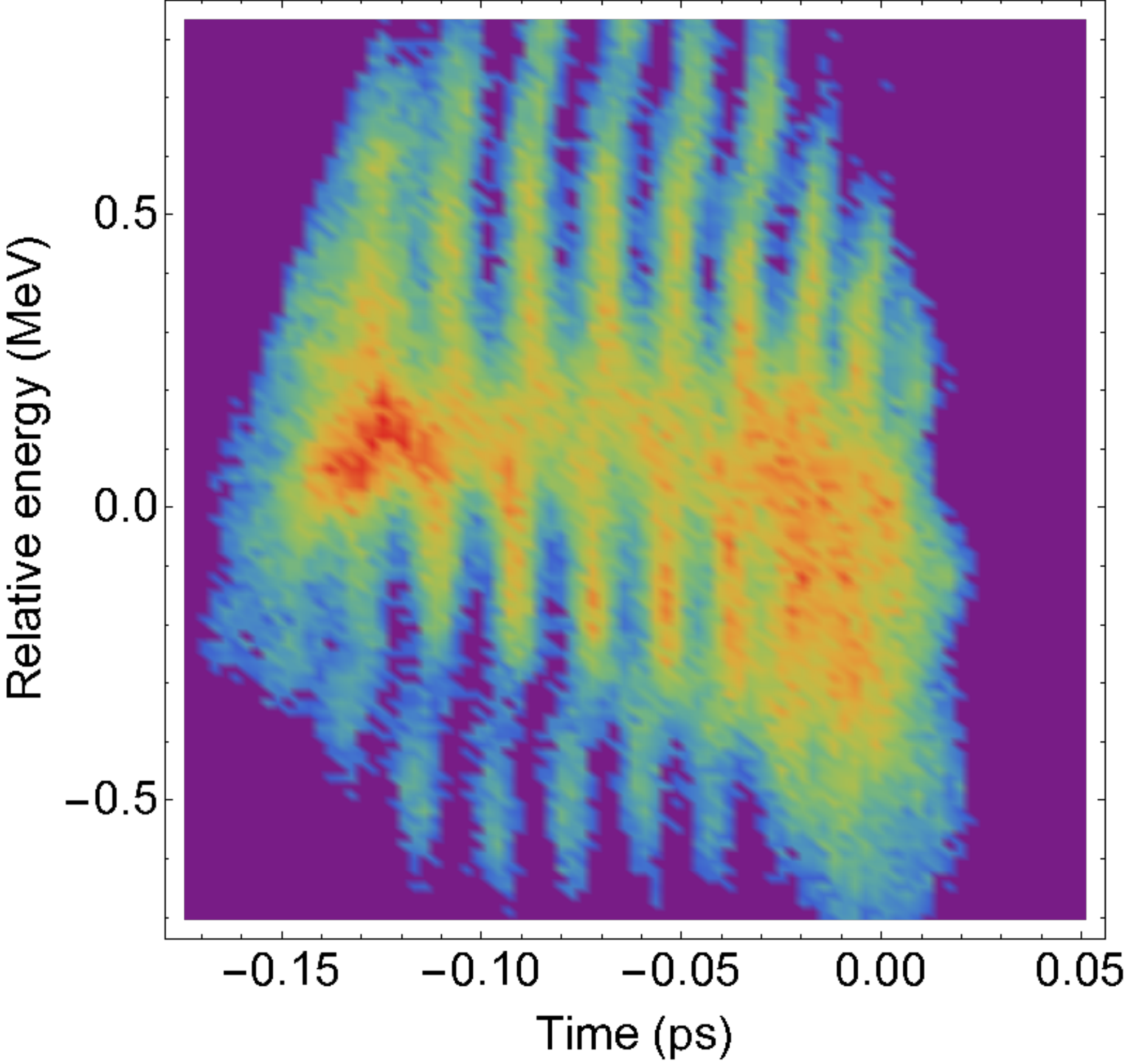} 			
			\label{fig:lpsbc1bc28501512ps}}
		\caption{Single-shot measured longitudinal phase space for BC1 and BC2 bending angles set to $105$\,\si{\milli\radian} and $28.5$\,\si{\milli\radian}, respectively, for a laser heater beating frequency of $1.8$\,\si{\tera\hertz}. The initial laser pulse energy from left to right was set to: $0.08$\,\si{\micro\joule}, $0.6$\,\si{\micro\joule}, $2.9$\,\si{\micro\joule} and $6.8$\,\si{\micro\joule}.} \label{fig:lpsbc1bc28512ps}
	\end{center}
\end{figure*}

We begin by analysing the features in the longitudinal phase space for the double compression scheme, i.e. using both magnetic chicanes. Some example images are shown in Fig.\,\ref{fig:lpsbc1bc28512ps}. In this case, the delay between the two laser pulses in the laser heater was set to $12$\,\si{\pico\second} ($\lambda_i = 166$\,\si{\micro\metre} and $\lambda_{f} = 7.2$\,\si{\micro\metre} for $C = 23$), and the initial laser pulse energy (before splitting and recombination) was varied between $0.1$\,\si{\micro\joule} and $10$\,\si{\micro\joule}, corresponding to an added energy spread (in the single pulse mode) between $5$\,\si{\kilo\electronvolt} and $50$\,\si{\kilo\electronvolt}. Qualitatively, it can be seen that, as the laser pulse energy increases, the longitudinal density modulations in the bunch become increasingly pronounced, and for the largest laser pulse energy shown (Fig.\,\ref{fig:lpsbc1bc28501512ps}), the modulation in the slice energy spread along the bunch becomes more prominent. These images are representative of the majority of measurements taken over $20$ shots, although there was some variation in the bunch length due to jitter in the RF structures. As a result, the statistical analysis presented below (Sec.\,\ref{sec:benchmarking}) was performed only for those bunches in which the current profile was close to the nominal value of around $600 \hbox{--} 650$\,\si{\ampere} in the bunch centroid.

%\onecolumngrid 

%\twocolumngrid

%\onecolumngrid

\begin{figure*}[bth!]
	\begin{center}
		\centering
		\subfloat[]{  
			\includegraphics[width=4.3cm]{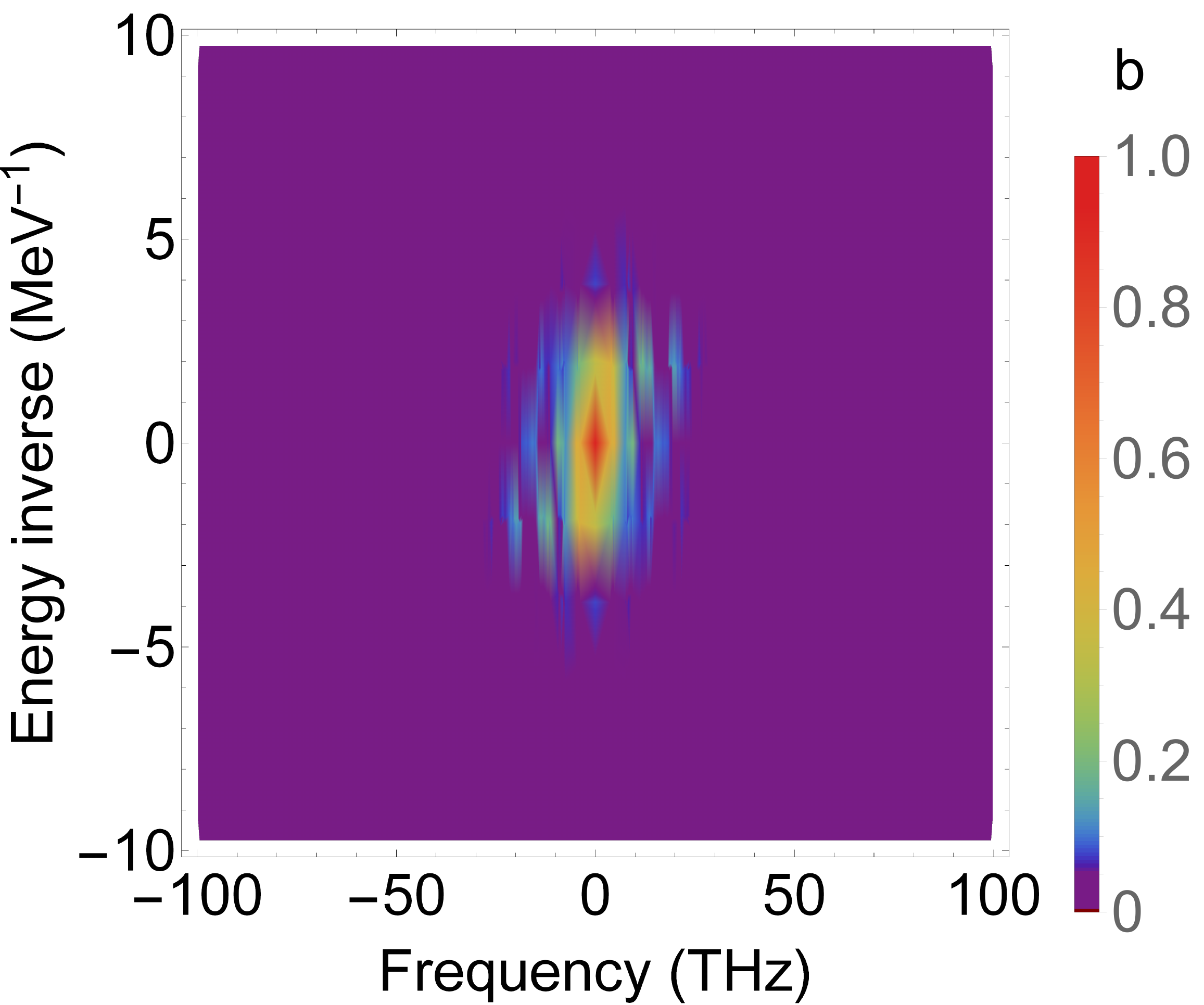} 
			\label{fig:ftbc1bc28500112ps}}
		\subfloat[]{  
			\includegraphics[width=4.3cm]{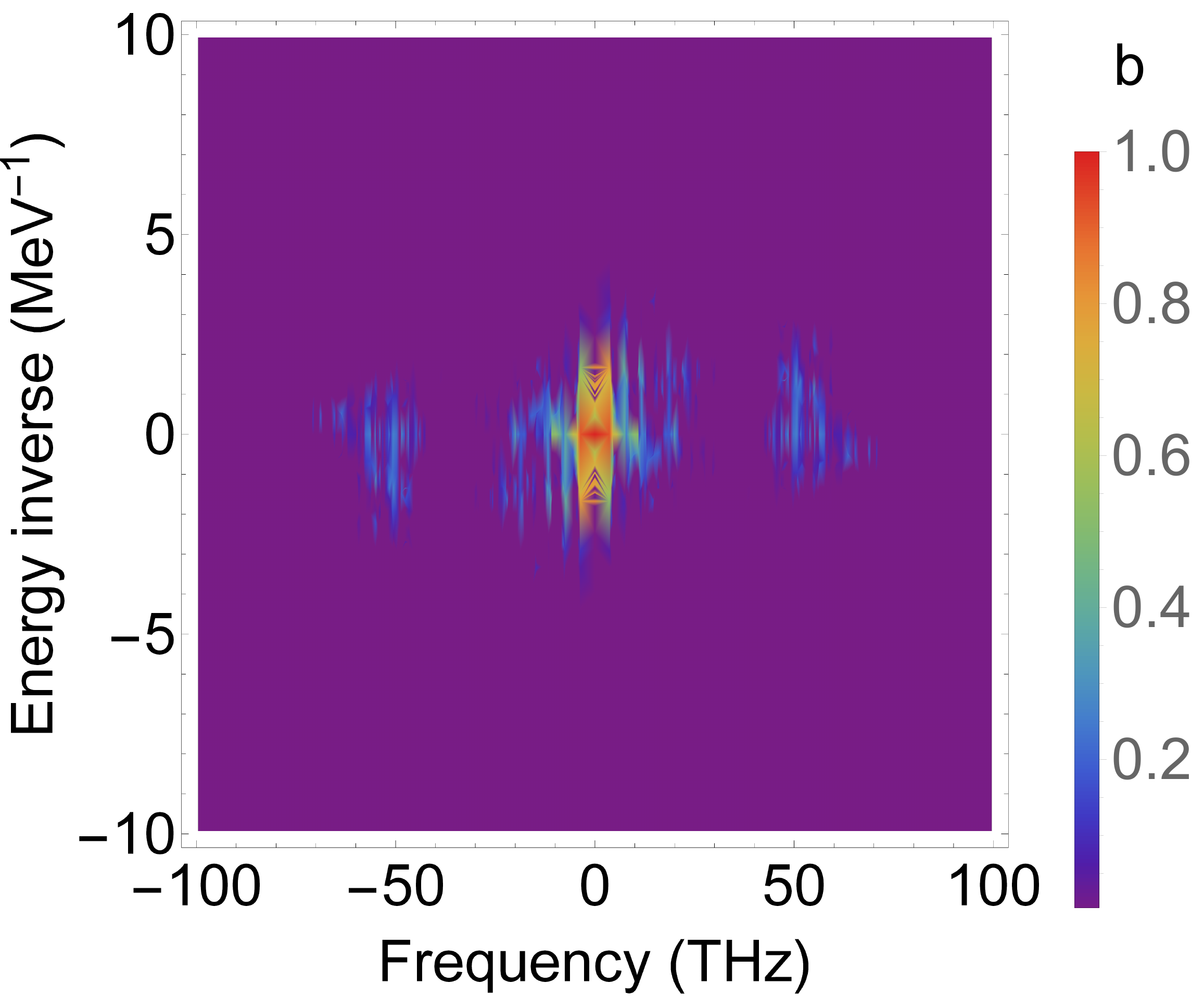} 
			\label{fig:ftbc1bc28505112ps}}
		\subfloat[]{  
			\includegraphics[width=4.3cm]{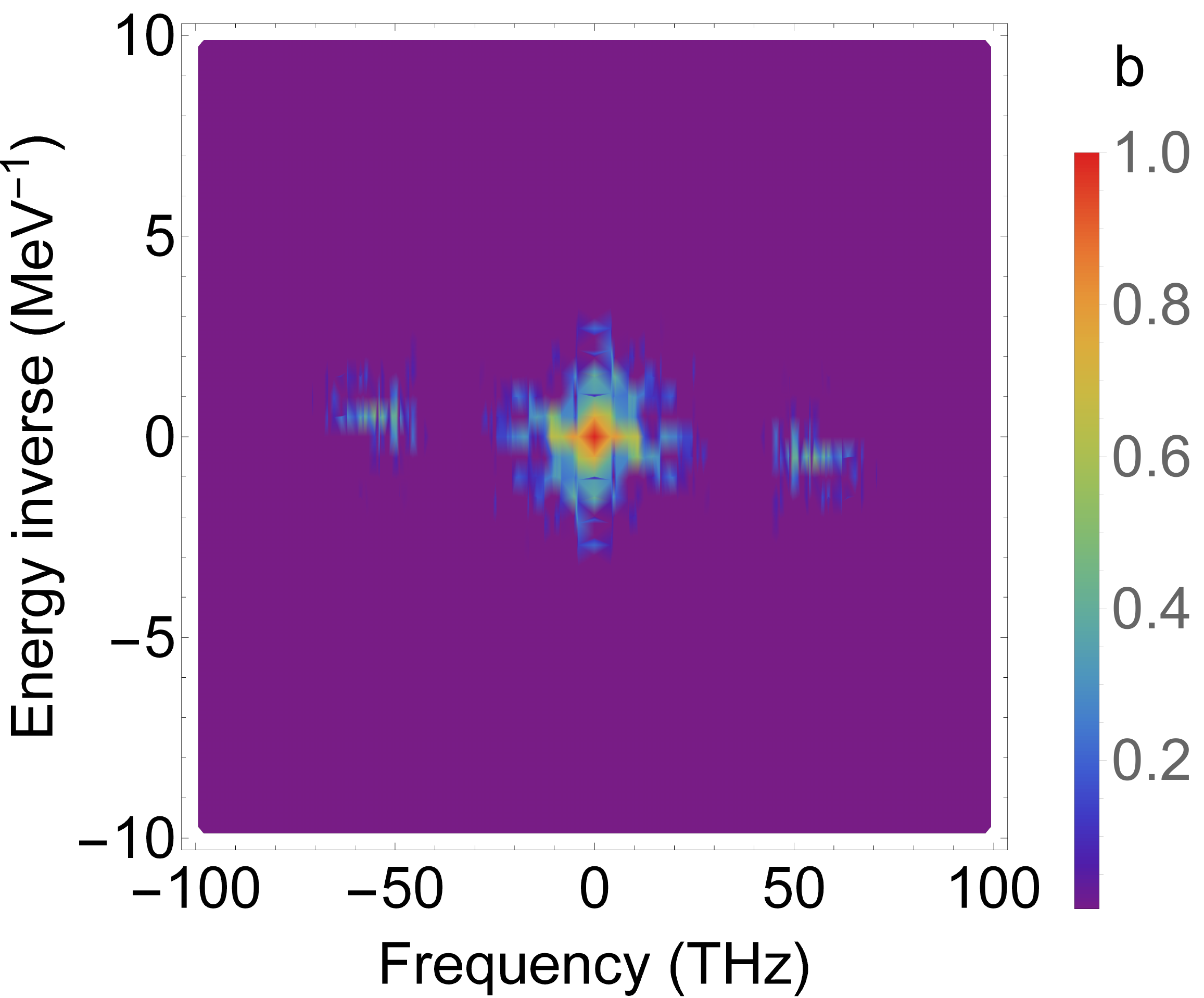} 
			\label{fig:ftbc1bc28501012ps}}
		\subfloat[]{  
			\includegraphics[width=4.3cm]{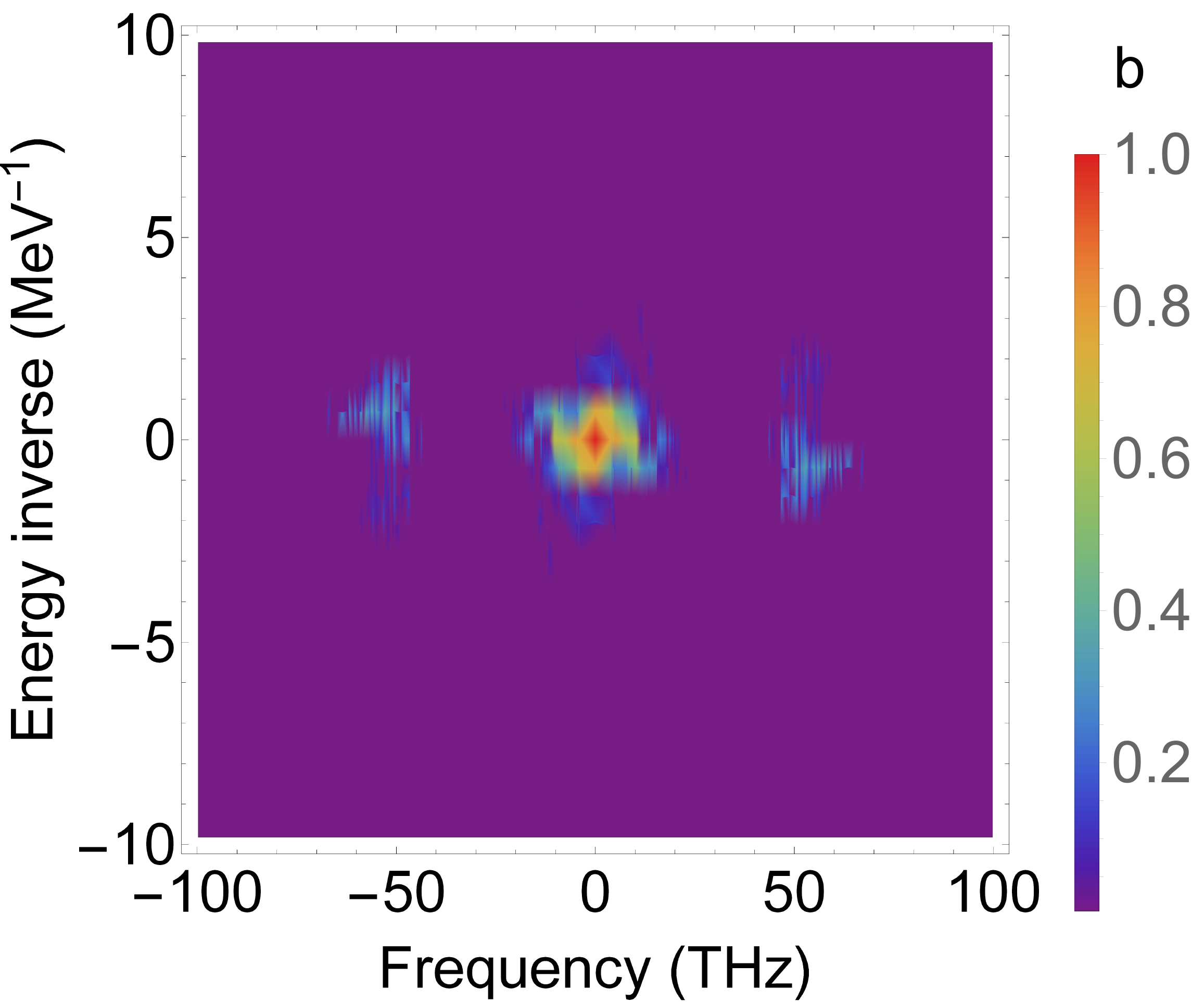} 			
			\label{fig:ftbc1bc28501512ps}}
		\caption{Measured Fourier transform (averaged over $20$ shots) for the same machine settings as in Fig.\,\ref{fig:lpsbc1bc28512ps} (BC1 + BC2 compression). The bunching factor $b$ is shown in the legend.} \label{fig:ftbc1bc28512ps}
	\end{center}
\end{figure*}

%\twocolumngrid

Through the use of 2D Fourier analysis, it is possible to measure the bunching factor along both axes of the longitudinal phase space simultaneously. This method has the added benefit of providing a measurement of the plasma oscillation phase of the microbunches as they transition between energy and density modulations \cite{SciRep.10.5059}. The variation in modulation frequency and bunching factor as a function of laser heater energy can be quantified by analysing the 2D Fourier transform of the longitudinal phase space. Examples of such transforms (corresponding to the machine settings used for the images in Fig.\,\ref{fig:lpsbc1bc28512ps}, but averaged over a number of shots) are shown in Fig.\,\ref{fig:ftbc1bc28512ps}. In this case, the sidebands that are offset from the DC term in the centre of each image in the Fourier transform characterise the bunching in energy and longitudinal density. The positions and amplitudes of the sidebands provide measurements of the microbunching period, phase and amplitude.

In the case of a $12$\,\si{\pico\second} initial beating delay, corresponding to an initial modulation frequency $\nu_i$ of $1.8$\,\si{\tera\hertz}, the final measured bunching period corresponds to a frequency $\nu_f$ in the range $40 \hbox{--} 45$\,\si{\tera\hertz}. By normalising the pixel intensity in Fourier space to the maximal value (at the centre), a measurement of the bunching factor as a function of frequency in both dimensions (energy and time) can be obtained (see Sec.\,\ref{subsec:bunching_factor} below). % This central term represents the bulk structure of the bunch, and it is all that remains if we apply a low-pass filter to the original longitudinal phase space images to artificially remove the modulations. 

\begin{figure}[bth!]
	\begin{center}
		\centering
		\subfloat[]{  
			\includegraphics[width=4.3cm]{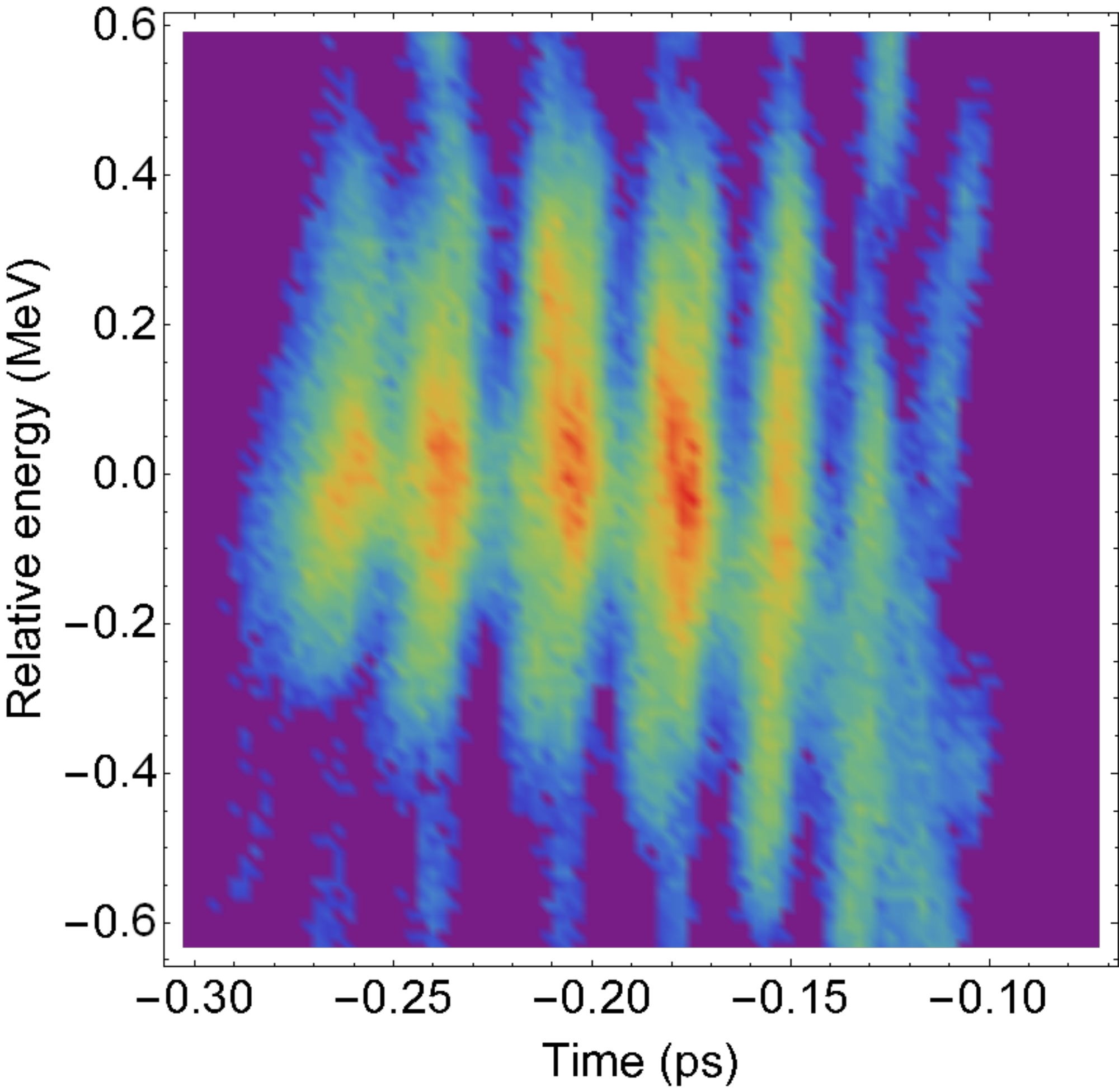} 
			\label{fig:lsbc1bc2850108ps}}
		\subfloat[]{  
			\includegraphics[width=4.3cm]{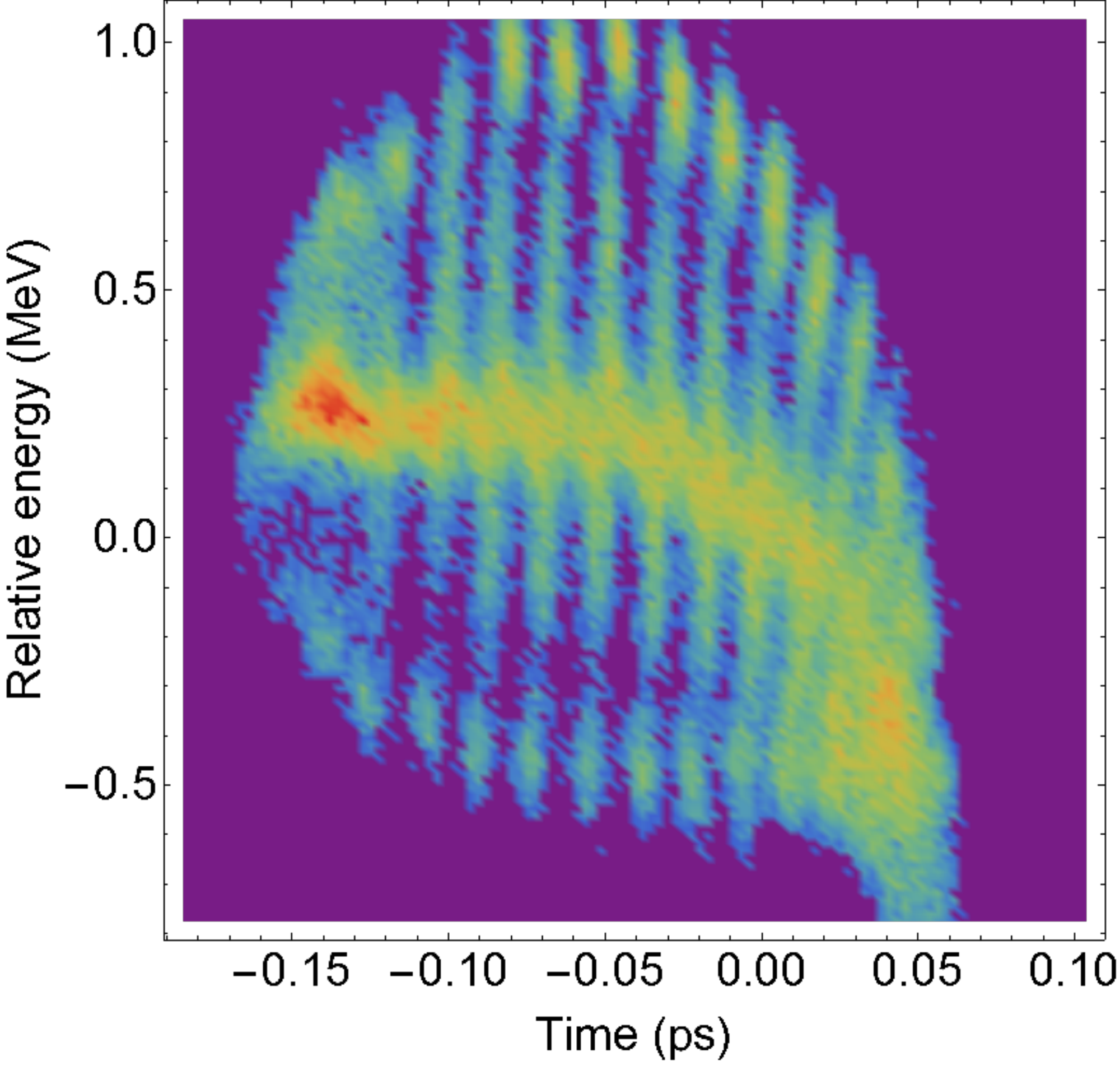} 
			\label{fig:lpsbc1bc28501016ps}}
		\vfill
		\subfloat[]{  
			\includegraphics[width=4.3cm]{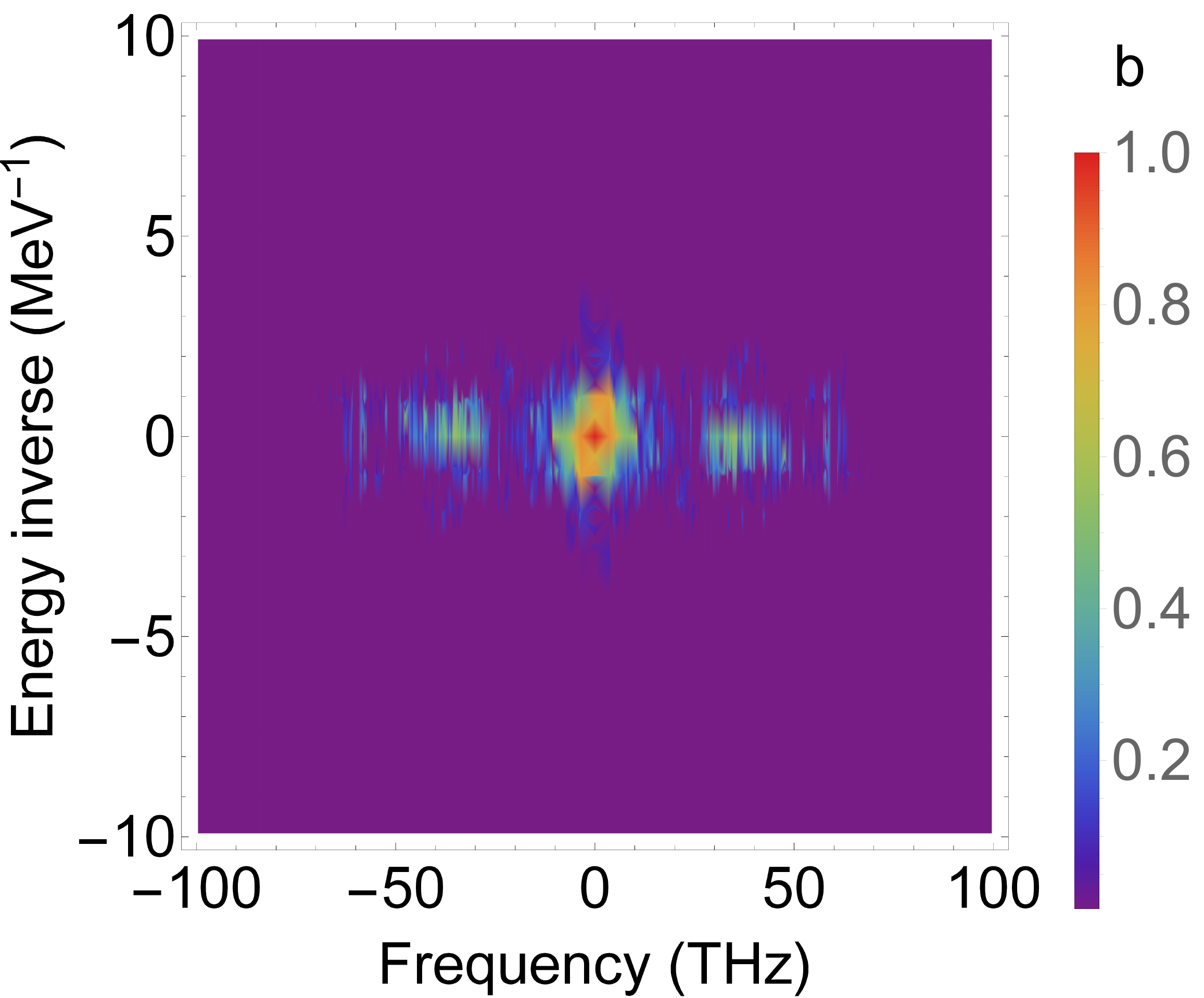} 
			\label{fig:ftbc1bc2850108ps}}
		\subfloat[]{  
			\includegraphics[width=4.3cm]{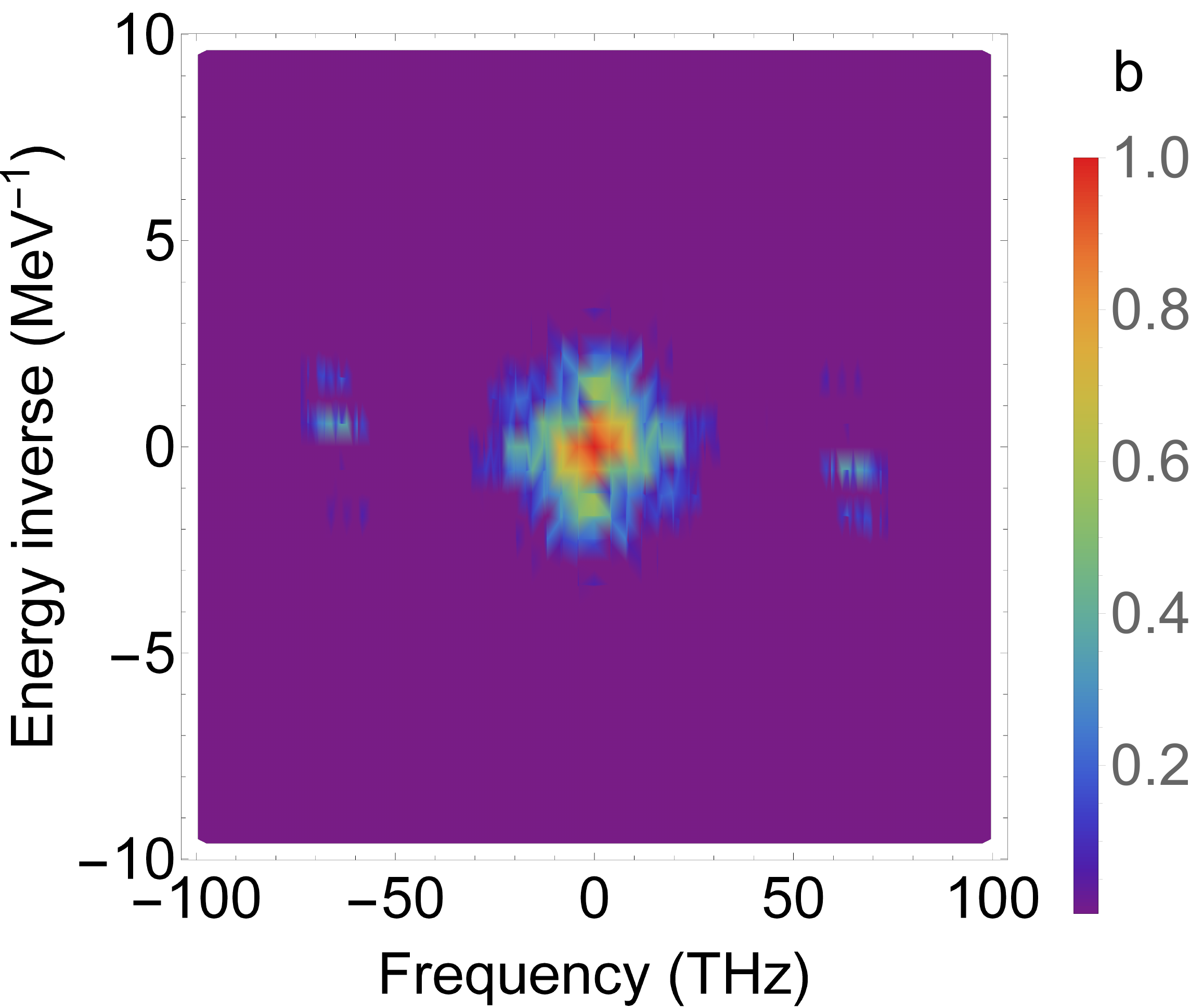} 			
			\label{fig:ftbc1bc28501016ps}}
		\caption{Single-shot measured longitudinal phase space (above) and Fourier transform (averaged over $20$ shots, below) for the same lattice settings as in Fig.\,\ref{fig:lpsbc1bc28512ps}, but with initial beating frequencies of $1.2$ (left) and $2.4$ (right) \si{\tera\hertz}, and an initial laser pulse energy of $3.5$\,\si{\micro\joule}. The colour scale, b, denotes the bunching factor (Eq.\,\ref{eq:bf}).} \label{fig:lpsft8ps16ps}
	\end{center}
\end{figure}

The relationship between the initial bunching frequency imposed by the modulated laser in the laser heater, the compression factor (around $23$) and the final bunching frequency has been measured for one laser heater beating frequency; we can now begin to analyse the development of microbunching for a range of different settings. The beating delay $\tau$ was varied between $8$ and $20$\,\si{\pico\second} for this compression scheme. Some example longitudinal phase space measurements for two settings of the initial beating frequency $\nu_i$ -- $1.2$\,\si{\tera\hertz} and $2.4$\,\si{\tera\hertz} -- are shown in the top row of plots in Fig.\,\ref{fig:lpsft8ps16ps}. In this case, the initial laser pulse energy was set to $3.5$\,\si{\micro\joule}. In the longitudinal phase spaces, it can be seen that the number of microbunches present in the bunch increases with the beating frequency. 

For a compression factor of $23$, these values of $\nu_i$ ($1.2$\,\si{\tera\hertz} and $2.4$\,\si{\tera\hertz}) correspond to a predicted final bunching frequency $\nu_f$ of $30$\,\si{\tera\hertz} and $55$\,\si{\tera\hertz}, respectively. We quantify this by measuring the position of the satellites in Fourier space relative to the DC term, as shown in the bottom row of plots in Fig.\,\ref{fig:lpsft8ps16ps}. The separation of the satellites from the DC term varies as a function of the initial modulation, and agreement between the predicted and measured final bunching frequency can clearly be seen. This relationship is characterised for all three compression schemes, discussed in Sec.\,\ref{subsec:modulation_period}. %This measurement is of particular interest, as the angular location of the satellites can reveal information about the plasma oscillation angle of the microbunches, and how this parameter varies for a given initial seeding modulation. 

%\onecolumngrid

\subsection{Single Compression}\label{subsec:single_compression}

A similar set of measurements to those detailed above were taken for bunches compressed using only BC2, the second bunch compressor. This resulted in the electron beam propagating with a lower peak current and uncompressed imposed modulations for a longer distance, meaning that short-range collective effects between the microbunches were relatively suppressed compared to a scheme in which the bunch is compressed at an earlier point in the lattice. In this machine configuration, a beam scraper was applied in the bunch compressor in order to suppress a large current spike at the head of the bunch (due to strong nonlinear compression), without having a measurable effect on the modulations in the rest of the bunch.

\begin{figure}[bth!]
	\begin{center}
		\centering
		\subfloat[]{
			\includegraphics[width=4.3cm]{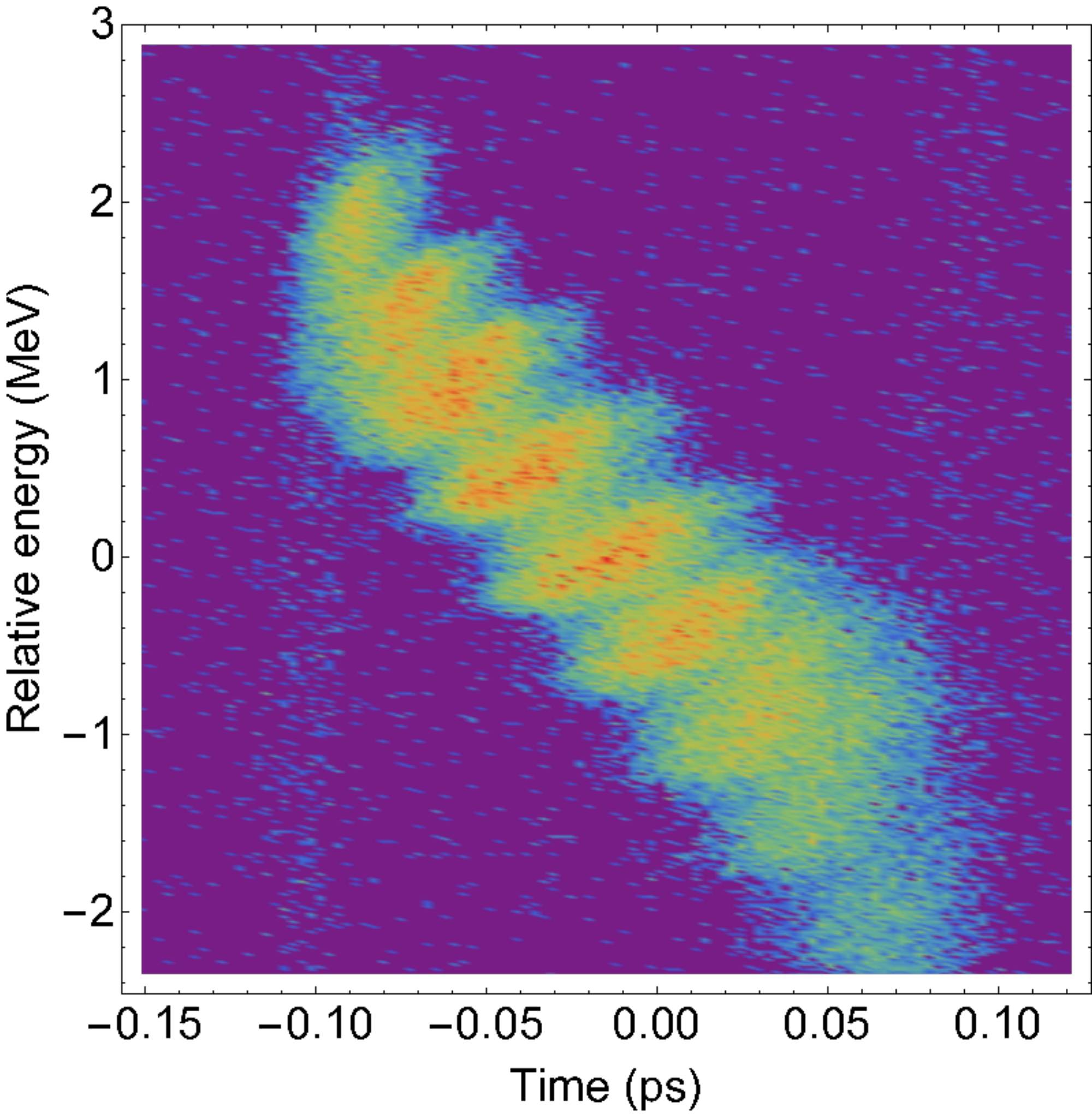} 
			\label{fig:lpsbc101512ps}}
		\subfloat[]{  
			\includegraphics[width=4.3cm]{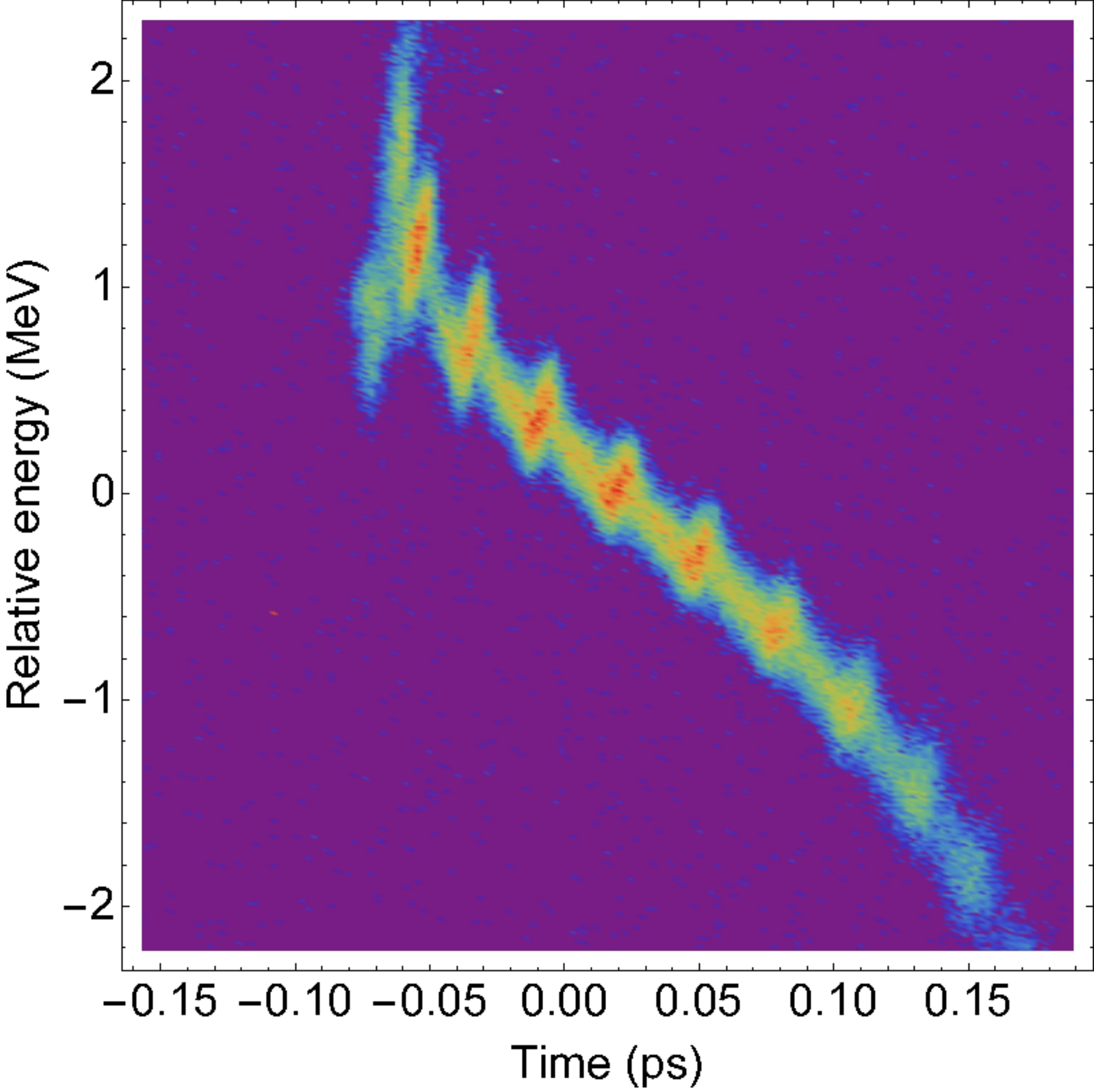} 			
			\label{fig:lpsbc201012ps}}
		\caption{Single-shot measured longitudinal phase space for bunches compressed using BC1-only (left) and BC2-only (right), with an initial beating frequency of $1.8$\si{\tera\hertz}.} \label{fig:lpsbc1onlybc2only}
	\end{center}
\end{figure}

In the case of bunches compressed using BC1 only, the longitudinal charge density of the bunch was high during the remainder of the acceleration process. This resulted in a larger impact on the energy spread of the bunch via Landau damping, and so the range of viable settings of the laser heater power was smaller than in the other two compression schemes: for $1$\,\si{\micro\joule} laser pulse energy and above, no significant modulation was observed on the bunch at the end of the linac. Example images of the measured longitudinal phase spaces for these two configurations -- with an initial beating frequency of $1.8$\,\si{\tera\hertz} -- are shown in Fig.\,\ref{fig:lpsbc1onlybc2only}. Across all three compression schemes, the final bunching periods observed generally agree well with the wavelength of the modulated laser heater pulse (see Sec.\,\ref{subsec:modulation_period} for further analysis), demonstrating the flexibility of this technique for producing strongly modulated bunches over a range of bunching periods. 

\section{Comparison Between Measurements and Simulations}\label{sec:benchmarking}

There are a number of parameters that can be extracted from both the measurements and simulations of the longitudinal phase space of modulated beams. In addition to the longitudinal phase space itself, microbunching parameters such as the modulation period on the bunch, the bunching factor, and the plasma oscillation phase can be measured using 2D Fourier analysis. In this section, these parameters will be characterised and compared between measurement and simulation. A wide range of initial modulation frequencies are imposed across multiple bunch compression schemes, and so we can study the development of the microbunching instability in detail.

\subsection{Simulated Longitudinal Phase Space Measurements}\label{subsec:lps_simulated}

\begin{figure*}[bth!]
	\begin{center}
		\centering
		\subfloat[]{  
			\includegraphics[width=4.3cm]{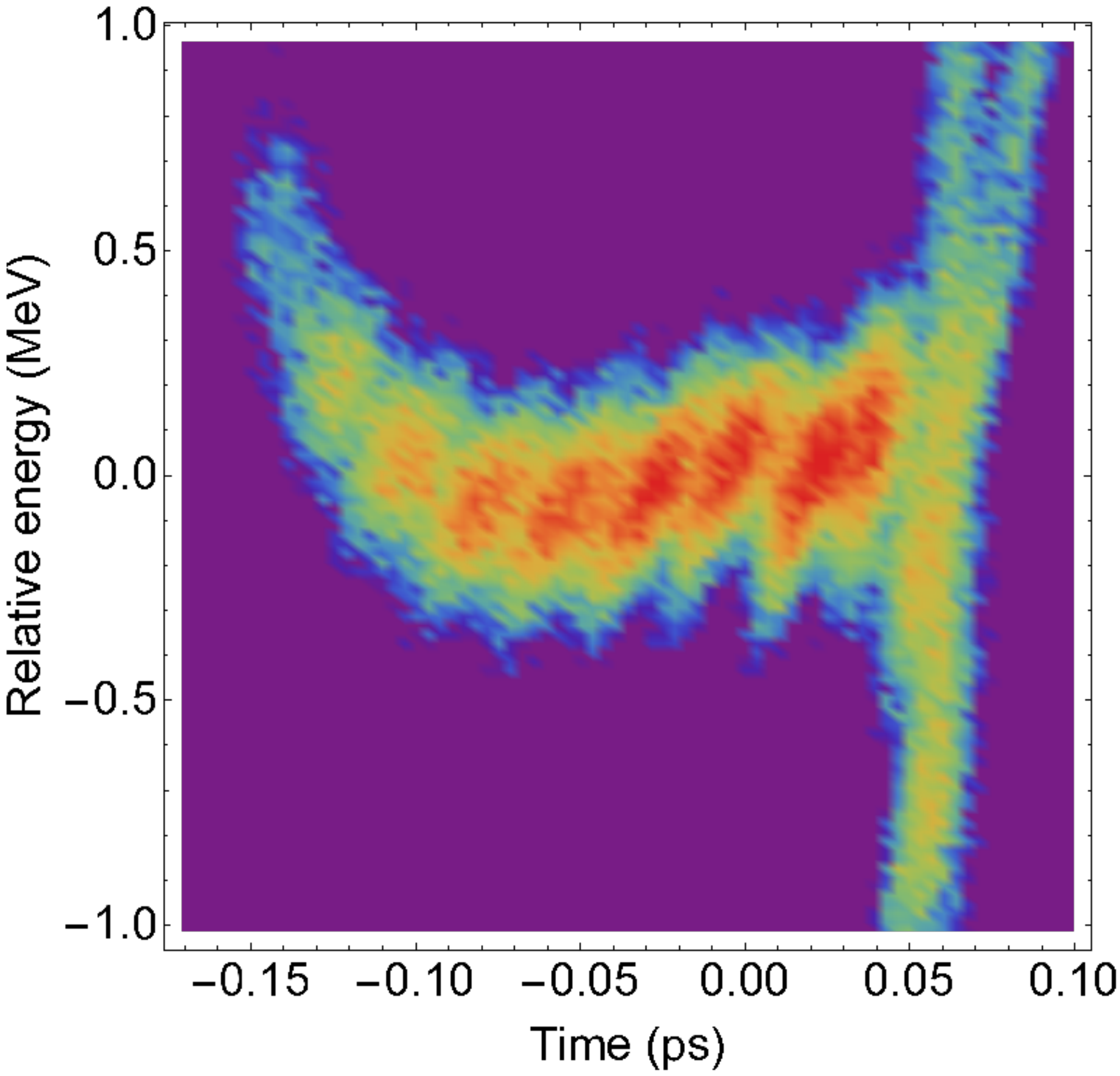} 
			\label{fig:lpsbc1bc28500112pssim}}
		\subfloat[]{  
			\includegraphics[width=4.3cm]{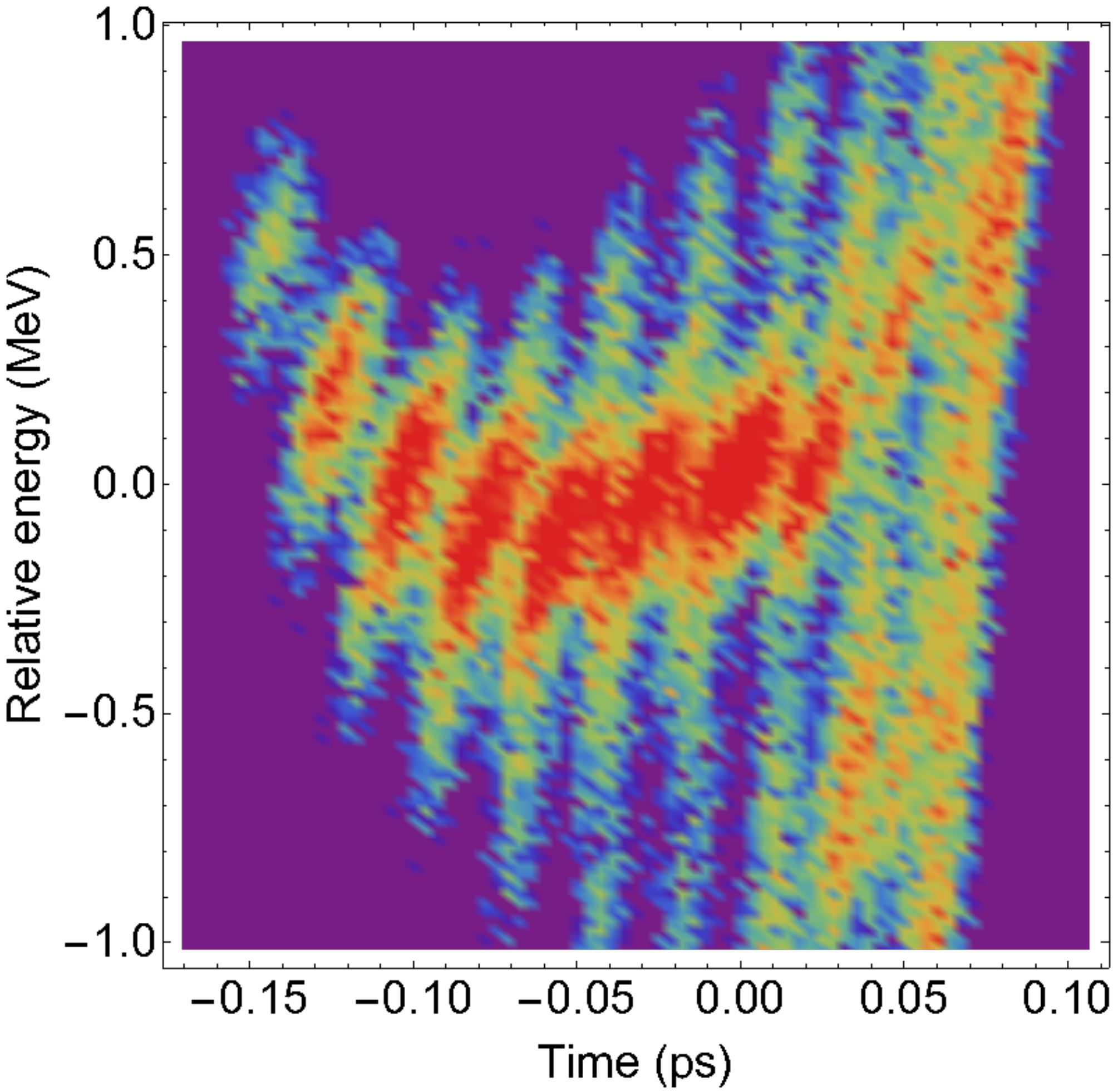} 
			\label{fig:lpsbc1bc28505112pssim}}
		\subfloat[]{  
			\includegraphics[width=4.3cm]{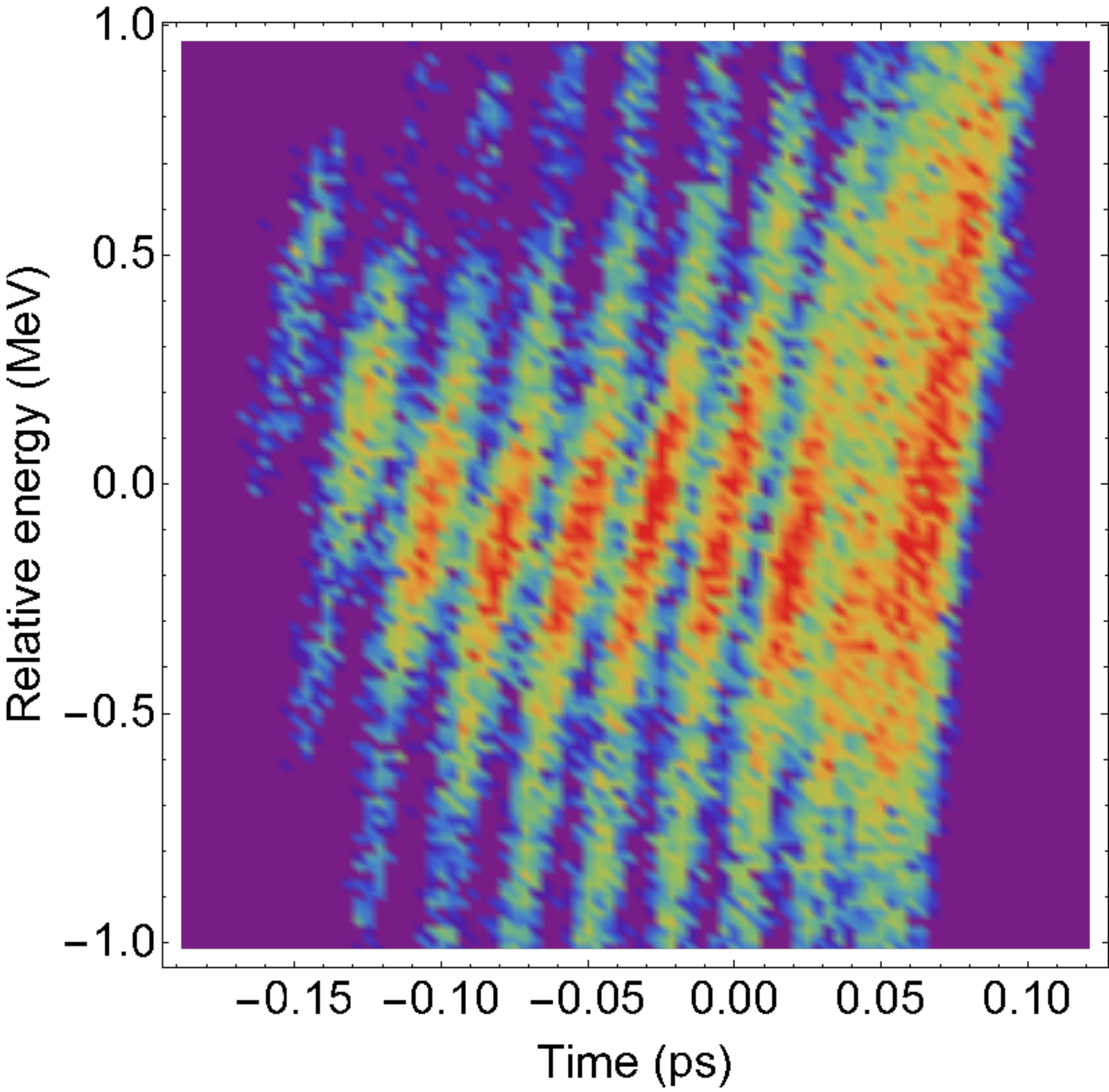} 
			\label{fig:lpsbc1bc28501012pssim}}
		\subfloat[]{  
			\includegraphics[width=4.3cm]{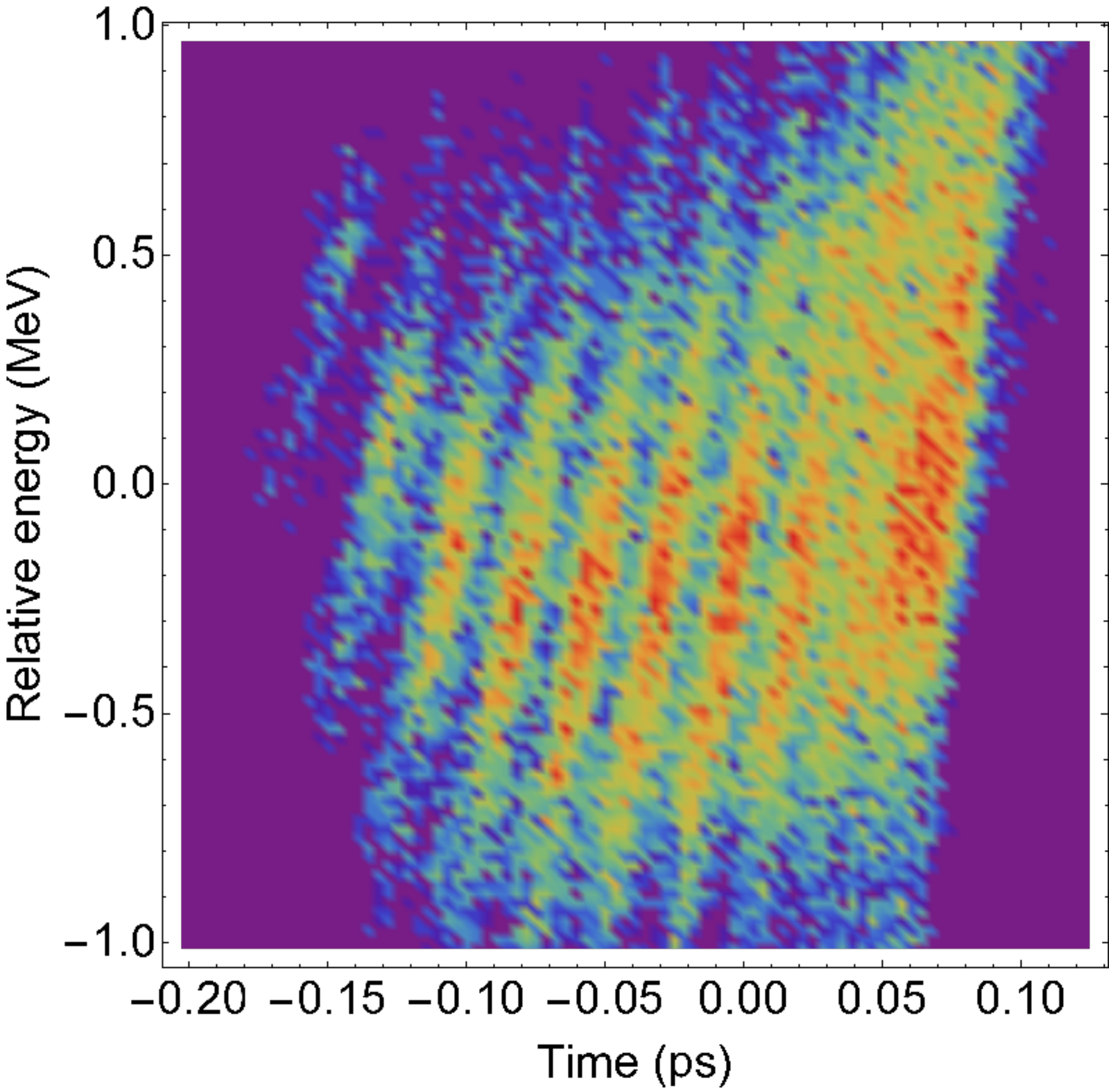} 			
			\label{fig:lpsbc1bc28501512pssim}}
		\caption{Simulated longitudinal phase spaces for the same machine settings as Fig.\,\ref{fig:lpsbc1bc28512ps} (BC1 + BC2 compression)} \label{fig:lpsbc1bc28512pssim}
	\end{center}
\end{figure*}

Having measured the effect of the modulated laser heater pulse on the electron beam for a range of pulse energies, it is instructive to compare these measurements with those from simulation. \textsc{Elegant} simulations were performed using the same parameters as were used in the measurements shown in Fig.\,\ref{fig:lpsbc1bc28512ps}. Collective effects and the modulated laser pulse in the laser heater were included (using the analytical expression for the laser pulse modulation, Eq.\ref{eq:lhbeating}). By including the effect of the deflecting cavity at the end of the linac, and using the same calibration factors as the measurement (pixel to \si{\milli\metre} to \si{\pico\second} and \si{\mega\electronvolt}), the simulation produced the longitudinal phase space images shown in Fig.\,\ref{fig:lpsbc1bc28512pssim}. 

%\twocolumngrid

Due to the large number of machine configurations, laser heater pulse energies and modulation wavelengths, the majority of simulations were run for only $10^5$ macroparticles -- a relatively small number compared with the real number of particles in the bunch. A convergence study was conducted for a subset of the experimental parameters using up to $10^7$ macroparticles, and varying the number of longitudinal density bins used for the CSR and LSC calculations. It was found that even this lower number of particles was able to reproduce the measured bunching. In cases where the modulation frequency on the bunch is smaller than that used in this experiment, it is expected that it would be necessary to run a larger simulation. Nevertheless, it can be seen that the simulation is able to reproduce the microbunching effects observed in the measured data: the final longitudinal phase space with a low-energy laser pulse exhibits similar macroscopic properties (in terms of total bunch length and energy spread); and the microbunching in the phase space becomes more prominent as the laser pulse energy increases. There is not an exact match between the simulated and experimentally measured distributions, in particular at the head and tail of the bunch, but the peak current in the bunch core and the periodicity of the modulations is similar (discussed below, Sec.\,\ref{subsec:modulation_period}). The slice energy spread for the BC2-only case in the bunch core is slightly larger than the measured value, but there is good agreement for the bunching factor between the two (see Sec.\,\ref{subsec:bunching_factor}) %The eventual reduction in modulation depth of the microbunches as the heating effect reaches its largest value is not observed to the same extent in the simulation, however. 

\begin{figure}[bth!]
	\begin{center}
		\centering
		\subfloat[]{
			\includegraphics[width=4.3cm]{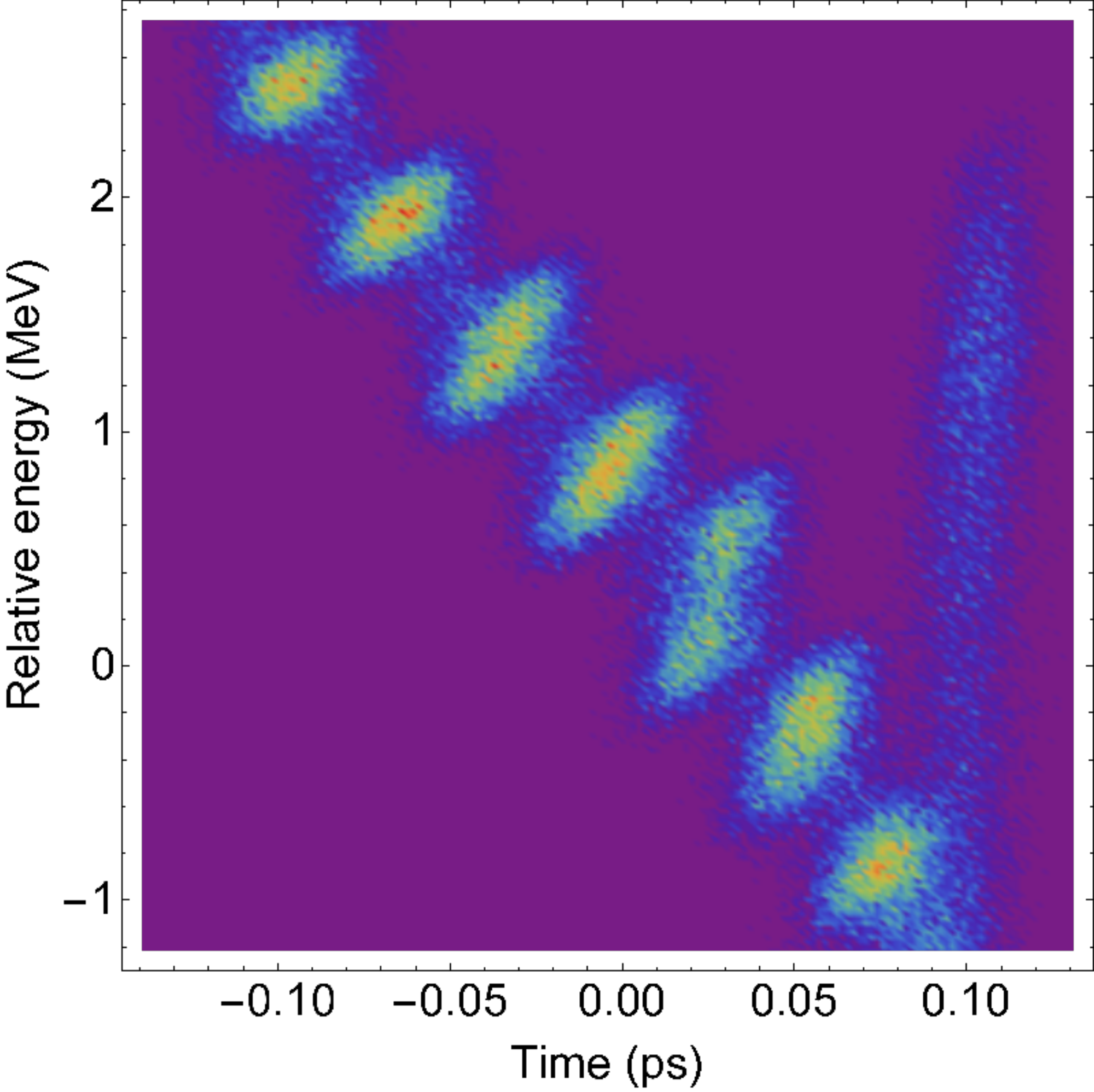} 
			\label{fig:lpsbc101512pssim}}
		\subfloat[]{  
			\includegraphics[width=4.3cm]{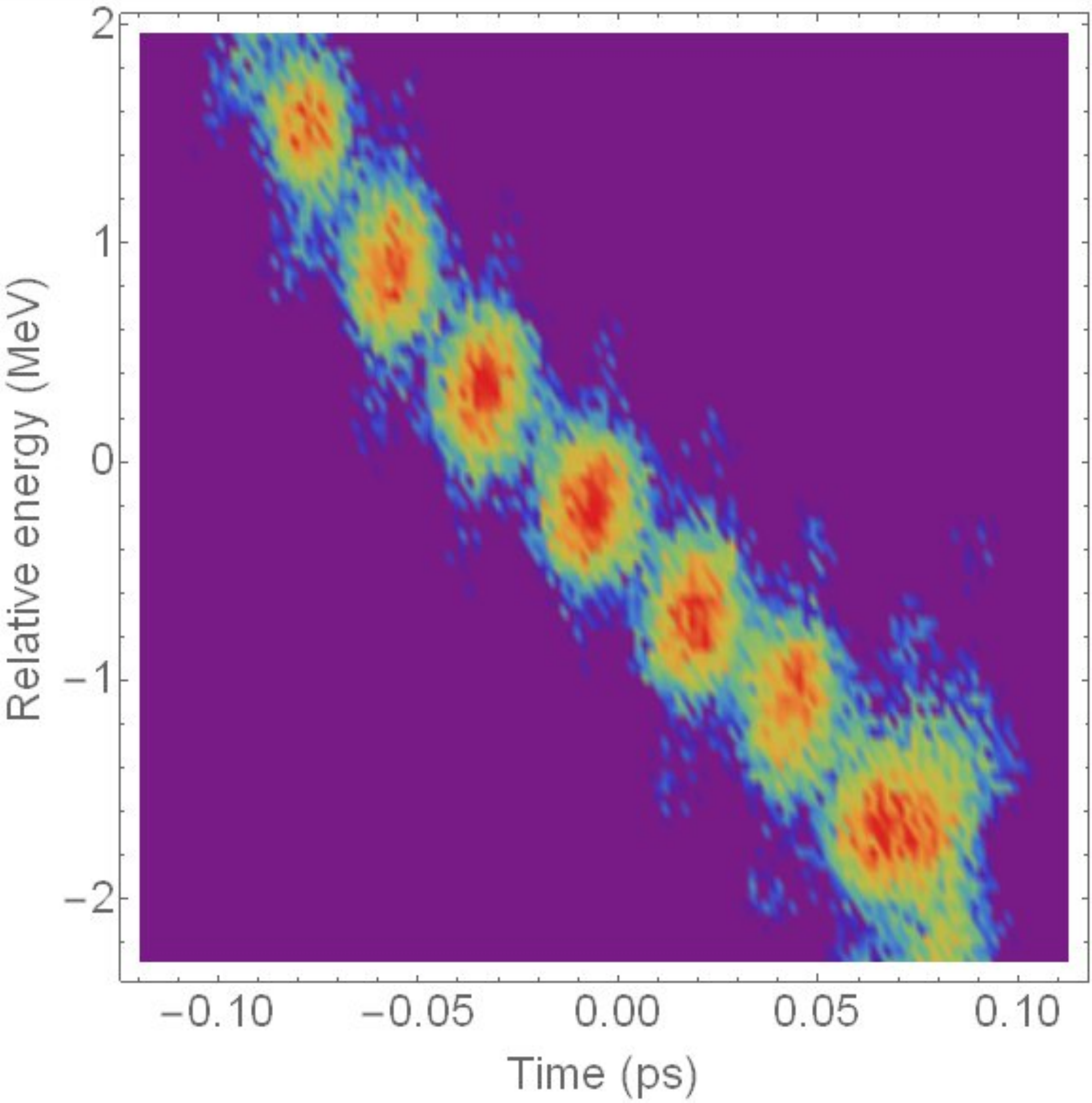} 			
			\label{fig:lpsbc201012pssim}}
		\caption{Single-shot simulated longitudinal phase space for bunches compressed using BC1-only (left) and BC2-only (right), with an initial beating frequency of $1.8$\si{\tera\hertz}.} \label{fig:lpsbc1onlybc2onlysim}
	\end{center}
\end{figure}

An example of the simulated longitudinal phase space for the two single-compression configurations -- with an initial beating frequency of $1.8$\,\si{\tera\hertz} -- is shown in Fig.\,\ref{fig:lpsbc1onlybc2onlysim}. It can be seen in both of these cases that the simulation is able to reproduce the experimentally measured macroscopic bunch properties (see Fig.\,\ref{fig:lpsbc1onlybc2only}), in terms of bunch length and energy spread, and that the modulations imposed have a similar periodicity. The bunching period and bunching factor is characterised in the subsections below.

\subsection{Modulation Period}\label{subsec:modulation_period}

The periodicity of microbunching in the electron bunch can be quantified as a function of either the longitudinal or energy density modulations. This can be achieved by projecting the 2D Fourier representation of the bunch images shown above onto the frequency or inverse energy axis. By doing this, we lose information concerning the correlative relationship between the microbunching along both axes, but this 1D information can provide a useful benchmark for simulation and theory. Additionally, one quantity may be more pertinent than another, for example when multi-colour FEL pulses are desired (in which case the bunching in energy is more significant), or for cases in which multiple bunches separated by time are required (and so longitudinal density modulations are important). 

In order to calculate the 1D projection of the bunching along the frequency axis, we select one of the satellites in Fourier space, and analyse only modulations above a specified frequency, so as not to include the DC term in our analysis, as this only provides information about the bulk structure of the bunch. As the laser heater pulse energy increases from zero, there is often a point at which the bunching factor at a given frequency reaches a maximum point, eventually decreasing in amplitude as the bunch slice energy spread becomes larger. This increase in slice energy spread reduces the depth of the modulations, in which case the modulated laser pulse effectively acts as a laser heater in its standard configuration. %The peak in the bunching observed at a period of $\lambda_f \approx 15$\,\si{\micro\metre} for the lowest setting of the laser heater energy is most likely due to the \lq natural\rq\,microbunching that occurs purely as a result of the collective effects in the machine, rather than being induced in the laser heater.

\begin{figure}[bth!]
	\begin{center}
		\includegraphics[width=8.6cm]{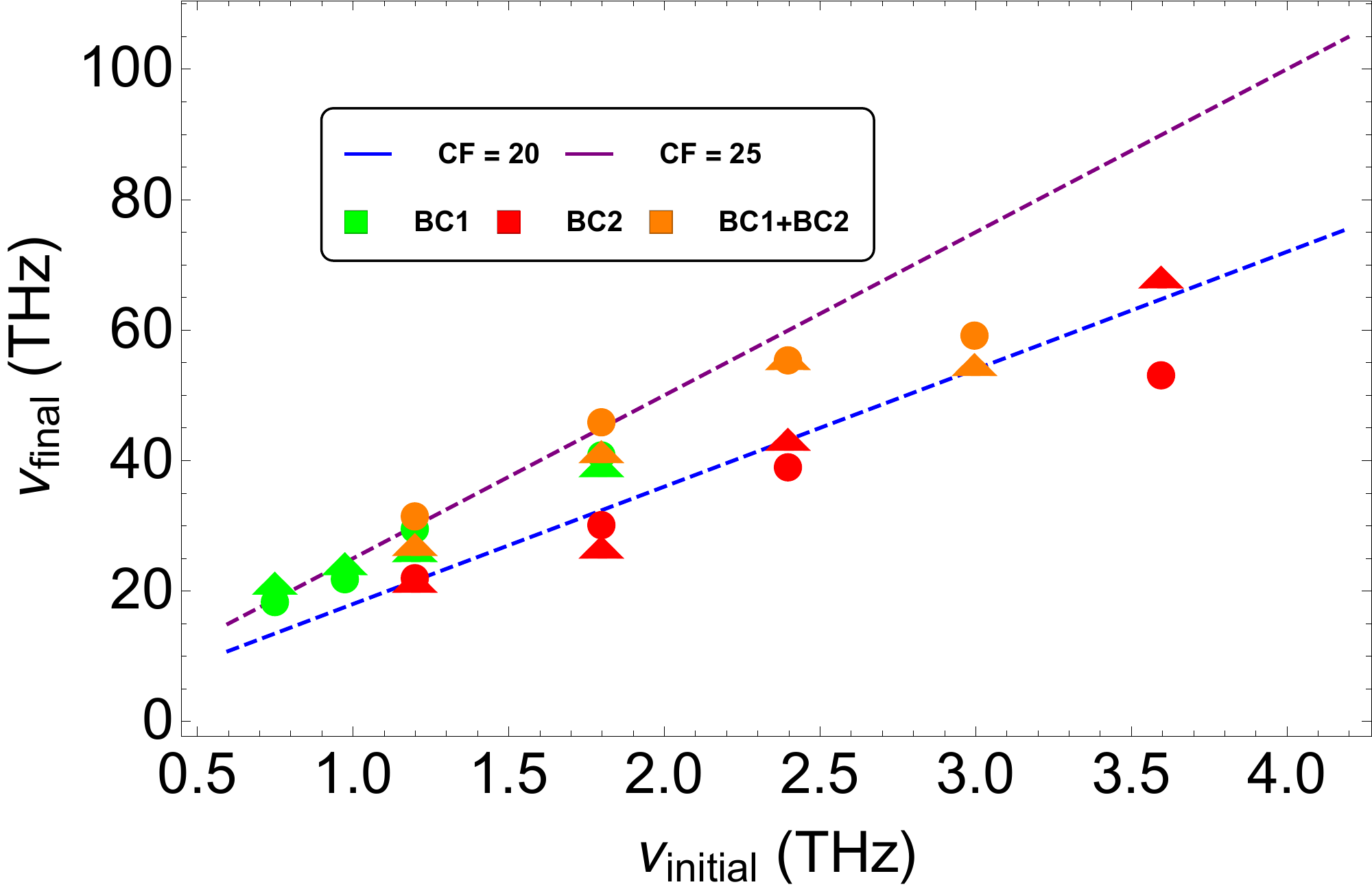}
		\caption{Final measured modulation frequency $\nu_{f}$ as a function of initial laser heater beating frequency $\nu_{i}$ for all three compression schemes. The dashed lines show a simple correlation between the initial and compressed laser frequency at two different compression factors ($CF \approx 20$ for BC2-only; $CF \approx 23 \hbox{--} 25$ for BC1-only and BC1+BC2); solid circles show measured values and solid triangles show simulated values.} \label{fig:beatingperiodall}
	\end{center}
\end{figure}

A full summary of the relationship between the initial beating frequency and the final measured modulation frequency for all three compression schemes is shown in Fig.\,\ref{fig:beatingperiodall}. By calculating the compressed frequency of the laser pulse in the laser heater, and comparing this with the measured frequency of the modulation on the bunch, it can be seen that, across all three compression schemes, the agreement is good. Each point on this plot represents the mean modulation frequency across a range of laser heater pulse energies -- this parameter stayed relatively constant for a wide range of values. \iffalse The average spectral bandwidth of the modulations remained at the level of a few \si{\tera\hertz} RMS in most cases.\fi A measurement of the compression factor that is independent of the modulation wavelength is the peak current of the bunch, which can be obtained by calculating the charge density in the bunch core. We found excellent agreement between these two measurements of the bunch compression factor.

Two different curves are presented for the compressed laser wavelength. This is due to the fact that the bunch compression was not exactly the same for all three lattice configurations; namely, the bunch was slightly longer in the BC2-only configuration (see Table\,\ref{table:fermi_parameters}). This presents a simple linear correlation between the initial laser beating frequency and its compressed value. There is some divergence between this simple model and the measured/simulated final modulation frequency on the bunch itself for larger initial beating frequency. This might be due to nonlinear effects impacting the modulation on the bunch, or the fact that the laser pulse energy was not sufficiently strong to imprint significant modulations on the bunch at these values of $\nu_i$. 

We also compare the measured values with those produced by simulation, and it can be seen that the agreement here is good in most cases. By varying either the initial beating frequency or the compression factor, this technique could easily produce modulations on the bunch over an even wider range. 

Due to the limitations of the pixel resolution of the measurement system, it was not possible to resolve microbunching at periods shorter than around $2$\,\si{\micro\metre}, as this approaches the Nyquist limit of the system. Only a small initial laser pulse energy ($< 2$\,\si{\micro\joule}) was required to override the \lq natural\rq\,microbunching such that the period at which the peak bunching factor is measured corresponds to the modulation imposed in the laser heater. It can also be seen that the simulated values of the bunching period agree well with the measurements.

\subsection{Bunching Factor}\label{subsec:bunching_factor}

\begin{figure}[bth!]
	\begin{center}
		\includegraphics[width=8.6cm]{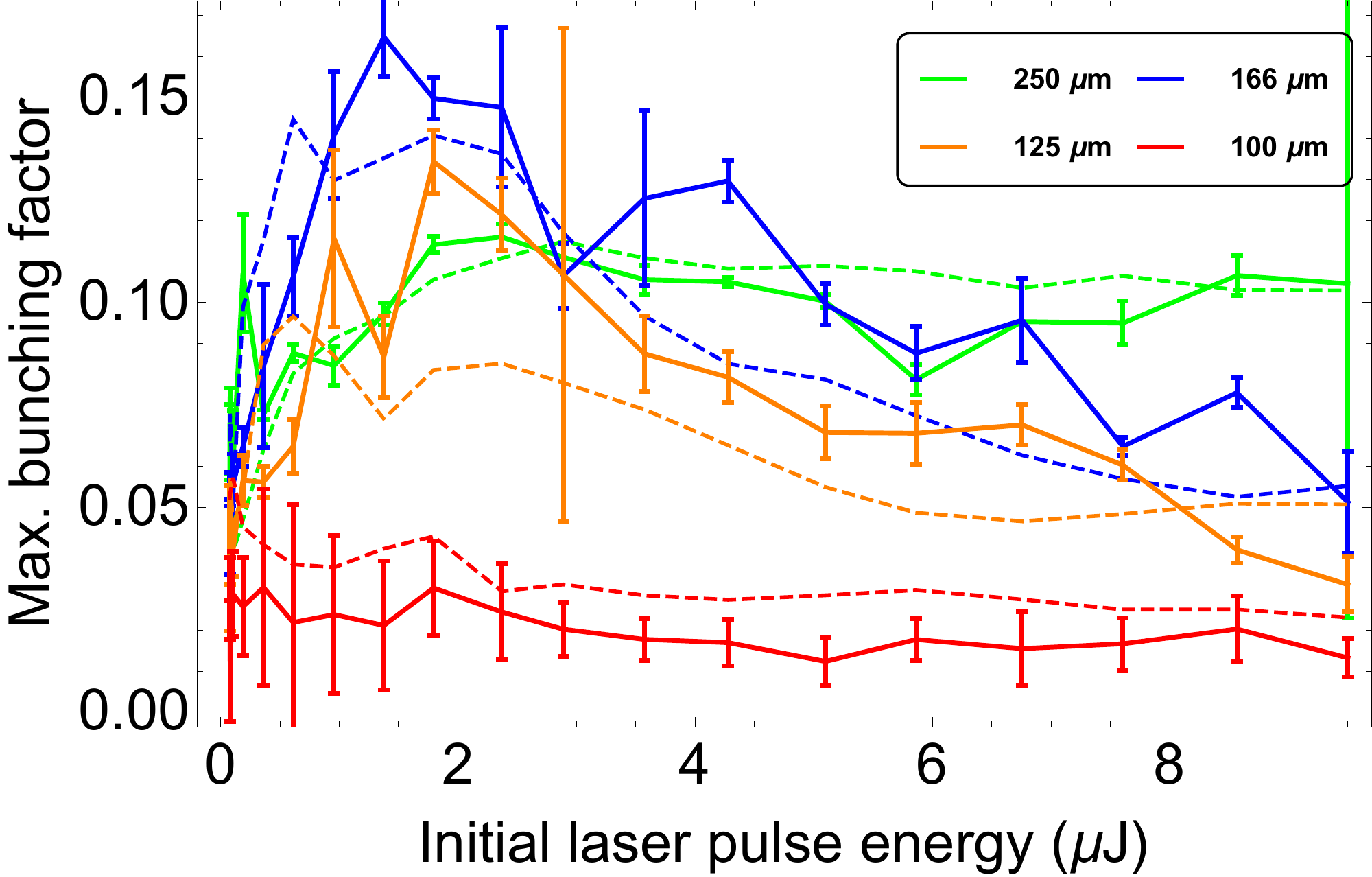}
		\caption{Maximum bunching factor as a function of initial laser pulse energy for a range of initial beating wavelengths in the double compression scheme. Measured bunching factor is shown by solid lines, and dashed lines show the simulated values. The error bars represent the standard deviation in maximum bunching factor over $20$ shots.} \label{fig:maxbfbc1bc285}
	\end{center}
\end{figure}

It is also possible to quantify the maximum bunching factor as a function of bunching period for each of the initial beating wavelength settings -- see Fig.\,\ref{fig:maxbfbc1bc285}. Each maximum corresponds to the largest bunching factor for a given laser pulse energy, irrespective of the modulation frequency. This plot shows the bunching factors for the double compression scheme. It can be seen that, in most cases, as the initial laser pulse energy increases, the maximum bunching measured reaches a peak for each setting of the initial beating frequency, eventually falling to a lower level as the energy spread induced in the laser heater undulator reduces the amplitude of the modulations in the bunch. We also note that the results from simulation are able to capture the trends observed across the full range of modulation wavelengths, although there are discrepancies in terms of the absolute values of the bunching factor in some cases. This is due to a combination of factors: the statistical variation in the experimental results arise both from the fact that microbunching is inherently a phenomenon based on noise, and from jitter in the RF system causing bunch length variation from shot to shot; additionally, the simulations do not account for the full 3D collective effects for the actual number of particles in the bunch, such as the influence of transverse space-charge and CSR forces.

\begin{figure}[bth!]
	\begin{center}
		\includegraphics[width=8.6cm]{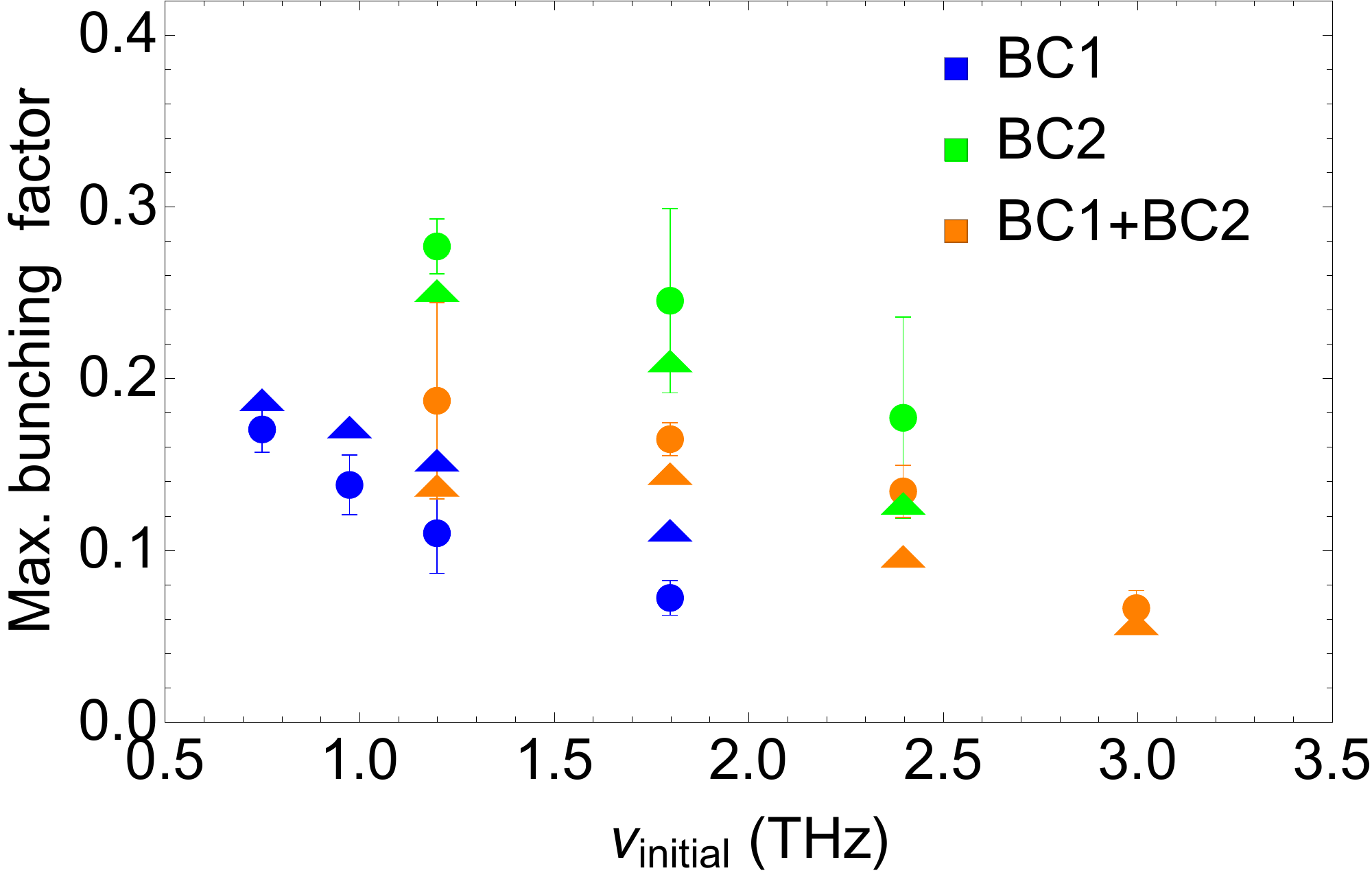}
		\caption{Maximum bunching factor -- measured (circles) and simulated (triangles) as a function of $\nu_i$ for the three compression schemes (given in the legend).} \label{fig:maxbfall}
	\end{center}
\end{figure}

A comparison between the simulated and measured maximum bunching factor across all three bunch compression schemes is shown in Fig.\,\ref{fig:maxbfall}. The results shown cover the ranges of initial laser beating frequencies used for each configuration. A general trend of higher levels of bunching at lower beating frequencies is observed (resulting in a longer bunching period on the bunch at the end of the linac). This is a result of the fact that the maximum intensity of the laser pulse for a given initial pulse energy decreases as the beating frequency increases (see Fig.\,\ref{fig:chirped_pulse_beating}). There is some discrepancy between measurement and simulation in terms of the absolute value of the maximum bunching factor in some cases, but our results show that the code is able to reproduce the trend observed experimentally.

\iffalse
Given the results of this analysis for one lattice configuration, we can now study the influence of the bunch compression factor across the same range of laser heater settings for a number of different BC2 angles, while maintaining the same compression in BC1. The direct proportionality between the initial and final modulation frequency -- related to each other via the bunch compression factor -- means that, as the $R_{56}$ parameter of both bunch compressors in tandem increases in magnitude, the maximal bunching factor will be found at increasingly shorter wavelengths. This is demonstrated in Fig.\,\textbf{FIGURE}, where the final modulation wavelength measured at the end of the linac is shown for a range of BC2 angles. In this case, the initial beating frequency was set to $1.8$\,\si{\tera\hertz}, and the initial laser pulse energy was set to $3.5$\,\si{\micro\joule}. As expected, the peak bunching wavelength decreases as the compression factor increases. Similar effects were observed for different settings of the laser beating frequency, although we again experience issues with resolving modulations below a certain wavelength. 
\fi

\subsection{Energy modulation}\label{subsec:energy_modulation}

\begin{figure}[bth!]
	\begin{center}
		\includegraphics[width=8.6cm]{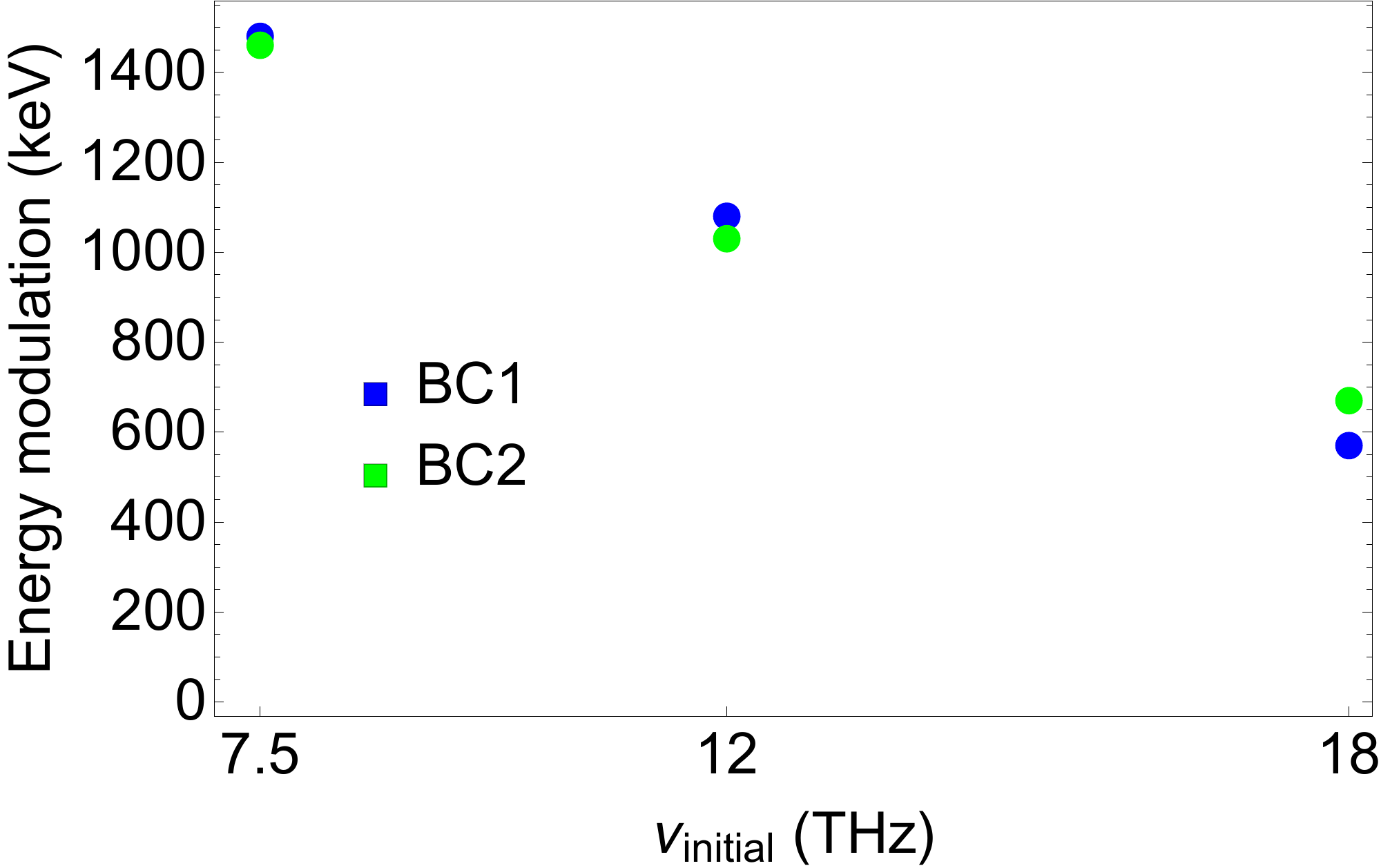}
		\caption{Measured energy modulation at the peak bunching factor as a function of initial beating frequencies for a fixed laser heater power. Results for BC1-only and BC2-only compression schemes are shown, since for the case of BC1+BC2 the bunches exhibited modulations purely in longitudinal density.} \label{fig:maxemodall}
	\end{center}
\end{figure}

As in the previous case of bunching in longitudinal density, the bunching in energy can also be studied (as shown in Fig.\,\ref{fig:maxemodall}). We restrict this analysis to just the single-compression schemes, which exhibited significant bunching in both energy and time, since in the case of double compression, the bunches were separated only in time, with the mean slice energy remaining constant. For the single compression schemes, we can project the 2D Fourier transform of the longitudinal phase space onto the energy axis, revealing the point of maximum bunching in energy. It can clearly be seen that, as the initial laser heater beating delay is increased (therefore increasing the frequency of modulations imposed on the bunch), there is a reduction in the measured periodicity along the energy axis. This bunching in energy is of interest for FELs, in particular schemes which aim to produce multi-colour pulses of light \cite{PhysRevLett.115.214801,NatComms.6.8486}. 

As with the analogous case for a projection onto the longitudinal axis described above (Sec.\,\ref{subsec:modulation_period}), the modulation in energy is prominent for low settings of the laser heater energy, and eventually decays as Landau damping reduces the depth of the modulations in the bunch. The results for the period and amplitude of bunching along the time and energy axes demonstrate the flexibility of the 2D Fourier transform technique for analysis of the longitudinal dynamics of modulated bunches.

\subsection{Microbunch angle}\label{subsec:microbunch_angle}

One feature of a microbunched beam that can only be parameterised by analysing the full longitudinal phase space is the angle of rotation of the microbunches. As mentioned above (Sec.\,\ref{sec:microbunching_instability}), the LSC forces within the bunch can cause a plasma oscillation between microbunching in energy and longitudinal density over a sufficiently long distance. This manifests itself as a rotation of the microbunches in longitudinal phase space \cite{SciRep.10.5059}. This process of rotation is further enhanced by bunch compression, which causes a shearing in the phase space. From the perspective of the generation of radiation in a light source, the plasma oscillation phase can be of vital importance for optimising the quality of the electron beam, and so its characterisation can provide valuable information. For example, in order to produce a monochromatic photon beam, the electron beam must also be monoenergetic, and this will not be the case if the plasma oscillation has caused bunching in energy at the undulator.

The plasma oscillation phase ($\theta_{p}$) of a microbunched beam can be measured using the 2D Fourier transform of the bunch image. As above, we focus our analysis on the satellites around the DC term in Fourier space. In this case, the plasma oscillation phase (i.e. the orientation angle of the microbunches) is given by the angle of the satellites in 2D frequency space with respect to the central term. Since the modulation period in energy and longitudinal density can be different by at least an order of magnitude (due to the difference between bunch length in \si{\pico\second} and the energy spread in \si{\mega\electronvolt}), we have measured the \lq normalised\rq\,plasma oscillation phase ($\theta_{p,N}$) as $\arctan\left(\bar{E}_{mod}/\bar{f}_{mod}\right)$, where $\bar{E}_{mod}, \bar{f}_{mod}$ denote the modulation frequencies in Fourier space for these two normalised axes, respectively. Using a method similar to transverse phase space tomography analysis \cite{NIMA.642.1.36}, the frequencies were normalised with respect to the bunch length and energy spread of the beam. In this case, a pure density modulation -- similar to that observed for the double compression scheme (see above, Sec.\,\ref{subsec:double_compression}) -- will give $\theta_{p,N} = 0$ or $\pi$, whereas a pure energy modulation produces $\theta_{p,N} = \pm \pi/2$. Any intermediate value between these pure modulations along one axis indicates that there is some mixing between bunching in energy and in density. More details on the measurement of this parameter can be found in Ref.\,\cite{SciRep.10.5059}.

\begin{figure}[bth!]
	\begin{center}
		\includegraphics[width=8.6cm]{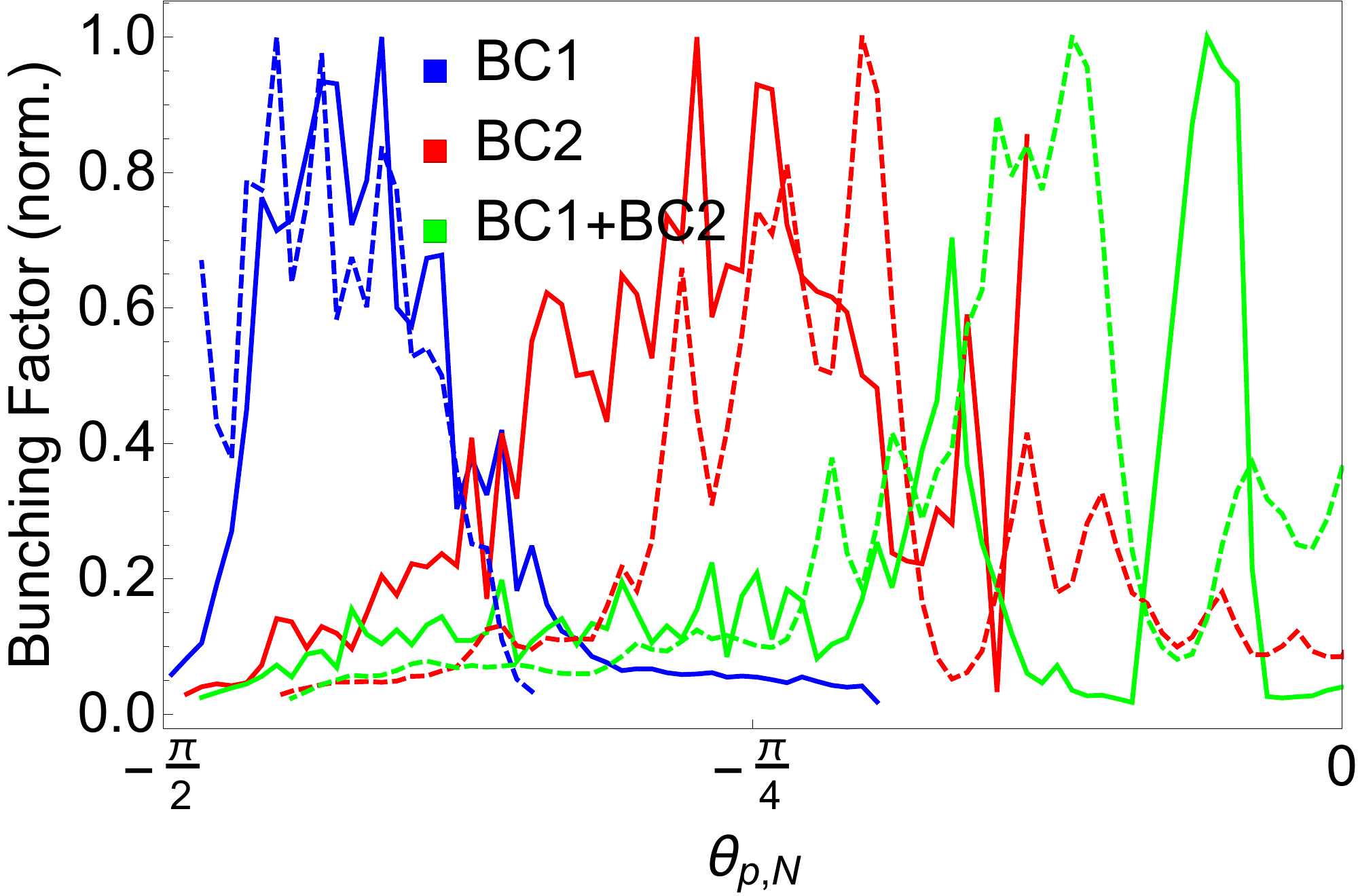}
		\caption{Bunching factor as a function of plasma oscillation angle $\theta_{p,N}$ for all three bunch compression schemes, each with an initial laser heater beating frequency of $1.8$\,\si{\tera\hertz}. Each curve represents the mean bunching factor over $20$ shots; solid lines show the measured bunching factor, and dashed lines show the corresponding simulations.} \label{fig:plasmaangleall}
	\end{center}
\end{figure}

Measurements of the bunching factor as a function of plasma oscillation phase are shown in Fig.\,\ref{fig:plasmaangleall} for all three compression schemes. In each case, the initial beating frequency was set to $1.8$\,\si{\tera\hertz}. We also show the simulated values for this parameter, with quite good agreement for all three compression schemes. %, but $\theta_{p,N}$ at the location of maximum bunching remained similar regardless of the initial modulation wavelength imposed on the bunch. This is because the plasma frequency depends only on the beam energy and current, and both of these parameters are fixed only by the compression scheme. 

It can be seen from the figure that, in the case of the double compression scheme, the bunch exhibited a modulation almost entirely in density (given that the maximum bunching factor is located around $\theta_{p,N} \approx 0$). This can be seen in the longitudinal phase space plots, in which the orientation of the microbunches results in a density modulation. In the case of single compression with BC1-only and BC2-only, there is a clear mixing between energy and density modulations, given that $\theta_{p,N}$ at the location of maximum bunching factor has a non-zero value. In such a case, the current profiles of these bunches exhibit clear modulations in both energy and longitudinal density, demonstrating that a purely 1D analysis would be insufficient to describe the microbunching for these more complex structures. 

\section{Conclusion}

In this paper, the development of the microbunching instability in a high-brightness electron linac has been investigated for a range of lattice and laser heater parameters. Through the use of a laser pulse that has been modulated by means of the chirped-pulse beating technique, and imposing this modulation on the energy spread of the electron bunch while propagating through the laser heater undulator, microbunching can be stimulated over a range of modulation wavelengths. By studying the longitudinal phase space of the electron bunch at the exit of the linac, in particular, by measuring the modulation amplitude (or bunching factor), it is possible to benchmark models for the microbunching instability. In addition, this technique provides a wide range of possible longitudinal phase space configurations through the variation of the laser pulse properties and the lattice parameters, and will be of interest for schemes which require microbunches that are separated in time or in energy. It was found that all three compression schemes resulted in similar maximum bunching factors. In the case of bunches compressed using only the first bunch compressor, the modulations on the bunch did not persist for even moderate settings of the laser heater energy, as a result of increased Landau damping for a bunch that propagated for a longer time with a short bunch length. On the other hand, the BC2-only compression scheme allowed for the manipulated bunches to persist with strong bunching over a wider range of laser heater energies. 

We used a high-fidelity simulation of the FERMI injector using the \textsc{GPT} code, based on experimentally measured photoinjector laser parameters and including collective effects such as 3D space charge and geometric wakefields in the RF cavities. The simulated injector bunch was then tracked through the remainder of the accelerator lattice using \textsc{Elegant}, a code that is well-suited to simulating the microbunching instability, as it can both tackle large numbers of particles relatively quickly and model the most pertinent collective effects (LSC and CSR). Given that the properties of the bunch longitudinal phase space are well defined at the exit of the injector and in the laser heater, we were able to produce a comparison between the simulated and experimental measurements of the bunching at the exit of the linac, analysing both the modulation depth and the final bunching period. In terms of the latter parameter, the simulation was able to reproduce accurately the measured results, demonstrating that this small-scale structure can persist through a long, complex accelerator lattice. The ELEGANT code employs 1D models for longitudinal space charge and coherent synchrotron radiation impedances, but it is still able to capture the physics of microbunching in its 2D content. This systematic comparison of microbunching in simulated and experimentally measured beams across a wide range of parameters suggests that the results from this code can be relied upon for future accelerator designs in which microbunching is expected to be an issue. The agreement between simulation and measurement for the bunching factor is quite good in many cases, and so this work provides a basis for simulating microbunching and longitudinal phase space manipulation in future accelerators, for example in beam-driven plasma wakefield acceleration or novel FEL schemes.

\section{Acknowledgements}
The authors acknowledge support from the Science \& Technology Facilities Council, UK, through a grant to the Cockcroft Institute, and the Industrial Liaison Office of Elettra Sincrotrone Trieste.
\bibliographystyle{unsrt}
\bibliography{references}

\end{document}